\documentclass[12pt]{article}
\usepackage{psfrag}
\usepackage{amsmath}
\usepackage{amsthm}
\usepackage{amssymb}
\usepackage{epsfig}
\usepackage{euscript}
\usepackage{array}
 \usepackage{cite}
\usepackage{cancel}
\usepackage{mathtools}
\usepackage{empheq}
\usepackage{graphicx}
\usepackage{subfigure}
\usepackage{upgreek}
\usepackage{epstopdf}
\usepackage{booktabs}
\usepackage{mathrsfs}
\usepackage{amsbsy}

\usepackage{hyperref}
\usepackage{verbatim}

\theoremstyle{definition}

\theoremstyle{remark}

\newcounter{multieqs}

\newcommand{\lab}[1]{\label{#1}}
\newcommand{\prt}[1]{{\left( {#1} \right)}}

\newcommand{\be}{\begin{equation}}
\newcommand{\ee}{\end{equation}}
\newcommand{\eq}[1]{(\ref{#1})}
\newcommand{\bit}{\begin{itemize}}  \newcommand{\eit}{\end{itemize}}
\newcommand{\ben}{\begin{enumerate}}  \newcommand{\een}{\end{enumerate}}

\newcommand{\bm}[1]{\mbox{\boldmath $#1$}}
\newcommand{\rf}[1]{(\ref{#1})}

\def\bd{\begin{document}}
\def\ed{\end{document}}
\def\bea{\begin{eqnarray}}
\def\eea{\end{eqnarray}}
\let\bm=\bibitem

\def\la{\langle}
\def\ra{\rangle}

\def\npb#1#2#3{Nucl. Phys. {\bf{B#1}} #3 (#2)}
\def\plb#1#2#3{Phys. Lett. {\bf{#1B}} #3 (#2)}
\def\prl#1#2#3{Phys. Rev. Lett. {\bf{#1}} #3 (#2)}
\def\prd#1#2#3{Phys. Rev. {D bf{#1}} #3 (#2)}
\def\cmp#1#2#3{Comm. Math. Phys. {\bf{#1}} #3 (#2)}
\def\cqg#1#2#3{Class. Quantum Grav. {\bf{#1}} #3 (#2)}
\def\nppsa#1#2#3{Nucl. Phys. B (Proc. Suppl.) {\bf{#1A}}#3 (#2)}
\def\ap#1#2#3{Ann. of Phys. {\bf{#1}} #3 (#2)}
\def\ijmp#1#2#3{Int. J. Mod. Phys. {\bf{A#1}} #3 (#2)}
\def\rmp#1#2#3{Rev. Mod. Phys. {\bf{#1}} #3 (#2)}
\def\mpla#1#2#3{Mod. Phys. Lett. {\bf A#1} #3 (#2)}
\def\jhep#1#2#3{J. High Energy Phys. {\bf #1} #3 (#2)}
\def\atmp#1#2#3{Adv. Theor. Math. Phys. {\bf #1} #3 (#2)}

\def\N{{\cal N}}
\def\sst{\scriptscriptstyle}
\def\thetabar{\bar\theta}
\def\Tr{{\rm Tr}}
\def\one{\mbox{1 \kern-.59em {\rm l}}}

\def\a{\alpha}      \def\da{{\dot\alpha}}  \def\dA{{\dot A}}
\def\b{\beta}       \def\db{{\dot\beta}}
\def\g{\gamma}  \def\G{\Gamma}  \def\dc{{\dot\gamma}}
\def\d{\delta}  \def\D{\Delta}  \def\ddt{\dot\delta}
\def\e{\epsilon}
\def\ve{\varepsilon}
\def\uve{\upvarepsilon}
\def\f{\phi}    \def\F{\Phi}    \def\vvf{\f}
\def\h{\eta}
\def\k{\kappa}
\def\l{\lambda} \def\L{\Lambda}
\def\m{\mu} \def\n{\nu}
\def\o{\omega}
\def\p{\pi} \def\P{\Pi}
\def\r{\rho}
\def\s{\sigma}  \def\S{\Sigma}
\def\t{\tau}
\def\th{\theta} \def\Th{\Theta} \def\vth{\vartheta}
\def\X{\Xeta}
\def\z{\zeta}

\def\na{\nabla}
\def\cA{{\mathscr A}} \def\cB{{\cal B}} \def\cC{{\cal C}}
\def\cD{{\cal D}} \def\cE{{\cal E}} \def\cF{{\cal F}}
\def\cG{{\cal G}} \def\cH{{\cal H}} \def\cI{{\cal I}}
\def\cJ{{\mathscr J}} \def\cK{{\cal K}} \def\cL{{\cal L}}
\def\cM{{\cal M}} \def\cN{{\cal N}} \def\cO{{\cal O}}
\def\cP{{\cal P}} \def\cQ{{\cal Q}} \def\cR{{\cal R}}
\def\cS{{\cal S}} \def\cT{{\cal T}} \def\cU{{\cal U}}
\def\cV{{\cal V}} \def\cW{{\cal W}} \def\cX{{\cal X}}
\def\cY{{\cal Y}} \def\cZ{{\cal Z}}

\def\ua{\underline{\alpha}}
\def\uc{\underline{\phantom{\alpha}}\!\!\!\gamma}
\def\um{\underline{\mu}}
\def\ud{\underline\delta}
\def\ue{\underline\epsilon}
\def\una{\underline a}\def\unA{\underline A}
\def\unb{\underline b}\def\unB{\underline B}
\def\unc{\underline c}\def\unC{\underline C}
\def\und{\underline d}\def\unD{\underline D}
\def\une{\underline e}\def\unE{\underline E}
\def\unf{\underline{\phantom{e}}\!\!\!\! f}\def\unF{\underline F}
\def\unm{\underline m}\def\unM{\underline M}
\def\unn{\underline n}\def\unN{\underline N}
\def\unp{\underline{\phantom{a}}\!\!\! p}\def\unP{\underline P}
\def\unq{\underline{\phantom{a}}\!\!\! q}
\def\unQ{\underline{\phantom{A}}\!\!\!\! Q}
\def\unH{\underline{H}}

\def\As {{A \hspace{-6.4pt} \slash}\;}
\def\bs {{b \hspace{-6.4pt} \slash}\;}
\def\Ds {{D \hspace{-6.4pt} \slash}\;}
\def\Gts {{\Gt \hspace{-6.4pt} \slash}\;}
\def\ds {{\del \hspace{-6.4pt} \slash}\;}
\def\ss {{\s \hspace{-6.4pt} \slash}\;}
\def\ks {{ k \hspace{-6.4pt} \slash}\;}
\def\ps {{p \hspace{-6.4pt} \slash}\;}
\def\xs {{x \hspace{-6.4pt} \slash}\;}
\def\pas {{{p_1} \hspace{-6.4pt} \slash}\;}
\def\pbs {{{p_2} \hspace{-6.4pt} \slash}\;}
\def\cFs {{{\cal F} \hspace{-6.4pt} \slash}\;}

\def\Ah{{\hat{A}}}
\def\Dh{{\hat{D}}}
\def\Gh{{\hat{G}}}
\def\Fh{{\hat{F}}}
\def\Ih{{\hat{I}}}
\def\Jh{{\hat{J}}}
\def\Kh{{\hat{K}}}
\def\Lh{{\hat{L}}}
\def\Ph{{\hat{P}}}
\def\Rh{{\hat{R}}}
\def\Vh{{\hat{V}}}
\def\Xh{{\hat{X}}}

\def\ah{{\hat{\a}}}
\def\bh{{\hat{\b}}}
\def\gh{{\hat{\g}}}
\def\dh{{\hat{\d}}}
\def\rh{{\hat{\r}}}
\def\hh{\hat{h}}
\def\uh{\hat{u}}
\def\xh{\hat{x}}
\def\yh{\hat{y}}
\def\ph{\hat{p}}
\def\xih{\hat{\xi}}
\def\chih{\hat{\chi}}
\def\Psih{\hat{\Psi}}
\def\phih{\hat{\phi}}

\def\psit{\tilde{\psi}}
\def\Psit{\tilde{\Psi}}
\def\Psibt{\tilde{\bar{Psi}}}

\def\st{\tilde{\sigma}}

\def\delt{\tilde{\delta}}
\def\Phit{\tilde{\Phi}}
\def\Phitb{\overline{\tilde{Phi}}}
\def\tht{\tilde{\th}}
\def\lt{\tilde{\l}}
\def\chit{\tilde{\chi}}
\def\phit{\tilde{\phi}}

\def\At{\tilde{A}}
\def\Bt{\tilde{B}}
\def\Ct{\tilde{C}}
\def\Dt{\tilde{D}}
\def\Et{\tilde{E}}
\def\Ft{\tilde{F}}
\def\Gt{\tilde{G}}
\def\Ht{\tilde{H}}
\def\It{\tilde{I}}
\def\Jt{\tilde{J}}
\def\Qt{\tilde{Q}}
\def\Rt{\tilde{R}}
\def\Mt{\tilde{M }}
\def\Nt{\tilde{N}}
\def\St{\tilde{S}}
\def\Vt{\tilde{V}}
\def\Xt{\tilde{X}}
\def\at{\tilde{a}}
\def\ct{\tilde{c}}
\def\dt{\tilde{d}}
\def\htt{\tilde{h}}
\def\ft{\tilde{f}}
\def\gt{\tilde{g}}
\def\pt{\tilde{p}}
\def\qt{\tilde{q}}
\def\vt{\tilde{v}}
\def\nt{\tilde{n}}
\def\ut{\tilde{u}}
\def\wt{\tilde{w}}
\def\zt{\tilde{z}}
\def\xt{\tilde{x}}
\def\yt{\tilde{y}}
\def\Psit{\tilde{\Psi}}
\def\vphit{\tilde{\varphi}}

\def\eb{\bar{\epsilon}}
\def\delb{\bar{\partial}}
\def\thb{\bar{\theta}}
\def\mub{\bar{\mu}}
\def\lamb{\bar{\l}}
\def\psib{\bar{\psi}}
\def\sb{\bar{\sigma}}
\def\xib{\bar{\xi}}
\def\chib{\bar{\chi}}

\def\Psib{\bar{\Psi}}
\def\Phib{\bar{\Phi}}
\def\Lamb{\bar{\Lambda}}
\def\Sb{{\overline \Sigma}}
\def\cb{\bar{c}}
\def\hb{\bar{h}}
\def\qb{\bar{q}}
\def\wb{\bar{w}}
\def\ub{\bar{u}}
\def\zb{{\bar{z}}}
\def\Hb{\bar{H}}
\def\Qb{{\bar Q}}
\def\Omegab{\overline{\Omega}}
\def\ob{\overline{\omega}}

\def\Ab{{\overline A}} \def\Bb{{\overline B}} \def\Cb{{\overline C}}
\def\Db{{\overline D}} \def\Eb{{\overline E}} \def\Fb{{\overline F}}
\def\Gb{{\overline G}}
\def\Ib{{\overline I}}
\def\Jb{{\overline J}} \def\Kb{{\overline K}} \def\Lb{{\overline L}}
\def\Mb{{\overline M}} \def\Nb{{\overline N}} \def\Ob{{\overline O}}
\def\Pb{{\overline P}}  \def\Rb{{\overline R}}
 \def\Tb{{\overline T}} \def\Ub{{\overline U}}
\def\Vb{{\overline V}} \def\Wb{{\overline W}} \def\Xb{{\overline X}}
\def\Yb{{\overline Y}} \def\Zb{{\overline Z}}

\def\fb{{\overline f}}
\def\gb{{\overline g}}
\def\mb{{\overline m}}
\def\lb{{\overline l}}
\def\yb{{\overline y}}

\def\ldel{{\overleftarrow{\del}}}
\def\rdel{{\overrightarrow{\del}}}
\def\ldeldel{{\overleftarrow{\del^2}}}
\def\rdeldel{{\overrightarrow{\del^2}}}
\def\ldelb{{\overleftarrow{\bar{\del}}}}
\def\rdelb{{\overrightarrow{\bar{\del}}}}
\def\ba{{\bf a}}
\def\bk{{\bf k}}
\def\bl{{\bf l}}
\def\bp{{\bf p}}
\def\bq{{\bf q}}
\def\br{{\bf r}}
\def\bt{{\bf t}}
\def\bu{{\bf u}}
\def\bv{{\bf v}}
\def\bx{{\bf x}}
\def\by{{\bf y}}
\def\bA{{\bf A}}
\def\bB{{\bf B}}
\def\bR{{\bf R}}
\def\bV{{\bf V}}

\def\bz{{\boldsymbol{\zeta}}}

\def\bone{{\bf 1}}

\def\va{{\vec a}}
\def\vk{{\vec k}}
\def\vp{{\vec p}}
\def\vq{{\vec q}}
\def\vx{{\vec x}}
\def\vy{{\vec y}}
\def\vu{{\vec u}}
\def\vv{{\vec v}}
\def \vH{{\vec H}}
\def \vg{{\vec g}}

\def\vs{{\vec \sigma}}
\def\vtau{{\vec \tau}}

\newcommand{\ov}[1]{\overrightarrow{#1}}

\def\frA{\mathfrak{A}}
\def\frB{\mathfrak{B}}
\def\frC{\mathfrak{C}}
\def\frD{\mathfrak{D}}
\def\frE{\mathfrak{E}}
\def\frF{\mathfrak{F}}
\def\frG{\mathfrak{G}}
\def\frH{\mathfrak{H}}
\def\frM{\mathfrak{M}}
\def\frN{\mathfrak{N}}
\def\frR{\mathfrak{R}}
\def\frW{\mathfrak{W}}

\def\fra{\mathfrak{a}}
\def\frb{\mathfrak{b}}
\def\frf{\mathfrak{f}}
\def\frg{\mathfrak{g}}
\def\frh{\mathfrak{h}}
\def\frl{\mathfrak{l}}
\def\frs{\mathfrak{s}}
\def\fri{\mathfrak{i}}
\def\frj{\mathfrak{j}}

\def\ma{\mathfrak{a}}
\def\mg{\mathfrak{g}}
\def\mh{\mathfrak{h}}
\def\mR{\mathfrak{R}}
\def\mN{\mathfrak{N}}

\newcommand{\nn}{{\nonumber}}

\def\d{\delta}\def\D{\Delta}\def\ddt{\dot\delta}

\def\pa{\partial} \def\del{\partial}
\def\xx{\times}
\def\uno{\mbox{1 \kern-.59em {\rm l}}}

\def\trp{^{\top}}
\def\inv{^{-1}}
\def\dag{\dagger}
\def\pr{^{\prime}}

\def\rar{\rightarrow}
\def\lar{\leftarrow}
\def\lrar{\leftrightarrow}

\newcommand{\0}{\,\!}      %this is just NOTHING!
\def\one{1\!\!1\,\,}
\def\im{\imath}
\def\jm{\jmath}

\newcommand{\tr}{\mbox{tr}}
\newcommand{\slsh}[1]{/ \!\!\!\! #1}

\def\vac{|0\rangle}
\def\lvac{\langle 0|}

\def\hlf{\frac{1}{2}}
\def\ove#1{\frac{1}{#1}}
\newcommand{\hot}[1]{\frac{#1}{2}}

\def\Box{\square}
\def\CC {\mathbb{C}}
\def\FF {\mathbb{F}}
\def\RR{\mathbb{R}}
\def\NN{\mathbb{N}}
\def\ZZ{\mathbb{Z}}
\def\bb#1{{\bf #1}}
\def\bcomment#1{}
\def\bfhat#1{{\bf \hat{#1}}}
\def\VEV#1{\left\langle #1\right\rangle}
\newcommand{\vevv}[1]{{\left< {#1} \right>}}

\newcommand{\ex}[1]{{\rm e}^{#1}} \def\ii{{\rm i}}

\newcommand{\lrbrk}[1]{\left(#1\right)}
\newcommand{\lrsbrk}[1]{\left[#1\right]}
\newcommand{\sfrac}[2]{{\textstyle\frac{#1}{#2}}}

\def\stw{{\sqrt{2}}}

\def\rf {{\rm f}}
\def\ri {{\rm i}}
\def\rj {{\rm j}}
\def\rn {{\rm n}}
\def\rk {{\rm k}}
\def\rl {{\rm l}}
\def\rs {{\scriptscriptstyle \rm S}}
\def\rt {{\scriptscriptstyle \rm T}}

\def\rQ {{\scriptscriptstyle \rm \cQ}}
\def\rR {{\scriptscriptstyle \rm \cR}}

\def\cQb{{\cal \Qb}}
\def\cRb{{\cal \Rb}}
\def\cWb{{\cal \Wb}}

\def\fd {{\rm N}}
\def\afd {{\overline{\rm N}}}

\def \II {I\hspace{-.1em}I\hspace{.1em}}
\def \IIA {\mbox{\II A\hspace{.2em}}}
\def \IIB {\mbox{\II B\hspace{.2em}}}
\def \gs {g^s}
\def \ls {\lambda^s}

\def \I {{\cal I}}
\def \qs {q\hspace{-.53em}/\hspace{.15em}}
\def \ks {k\hspace{-.53em}/\hspace{.15em}}
\def \YM {{\mbox{\tiny YM}}}
\def \gym {g_{\YM}}

\def \Lc {\L_c}
\def\IR{\relax{\rm I\kern-.18em R}}
\def \id {{\bf 1}}

\def\cci{\ell}
\def\ccj{\ell'}

\def\bbq{\pmb{q}}

\begin{document}
\begin{titlepage}
\begin{flushright}
%c3
\hfill{NCTS-TH/1607}
 \end{flushright}
\hfill
% \vskip 0.4in

 \begin{center}

%%%%%%%%%%%%%%%%%%%%%%%%%%%%%%%%%%%%%%%%%%%%%%%%%%
{\Large \bf The Thermal Bath of de Sitter from Holography
}\\[10mm]
%%%%%%%%%%%%%%%%%%%%%%%%%%%%%%%%%%%%%%%%%%%%%%%%%%

{\bf Chong-Sun Chu${}^{1,2}$,  Dimitrios
Giataganas${}^1$}

{\itshape ${}^1$ Physics Division, National Center for Theoretical
  Sciences, \\
 National Tsing-Hua University, Hsinchu, 30013, Taiwan}\\[1mm]
{\itshape ${}^2$ Department of Physics, National Tsing-Hua
  University,  Hsinchu 30013, Taiwan}

%d1
{\small \sffamily
cschu@phys.nthu.edu.tw~, dgiataganas@phys.cts.nthu.edu.tw
\\
}
\end{center}

\date{\today}

\begin{abstract}

We consider the AdS/dS CFT correspondence and study the nature of the
thermal bath of the de Sitter field theory using holography.
Unlike the temperature of a thermal field theory in flat spacetime,
the temperature of a superconformal field theory on de Sitter space
is an integral part of the theory and leaves intact
the conformal symmetry and supersymmetry. In the dual AdS side,
there is no  black hole.
%c3
Instead we have cosmological expansion of the de Sitter factor.
We consider a number of different observables,
such as the entanglement entropy, two point correlation function, %dg6
Wilson loops corresponding to static and spinning mesons in
the field theory, and study their thermal properties using holography.
The former two quantities have %dg6
trivial temperature dependence due to conformal
symmetry. We
compute the energy of the quark anti-quark bound state for a static %dg3
meson, as well as  the energy and  the
angular momentum for a spinning meson.  We find that
there is a maximum distance, as well as a maximum spin for the latter
case,
beyond which the bound state become unstable.
The temperature behavior of the physical quantities
in these meson systems are similar to that of
the usual thermal field theory
%c3
with holographic black hole dual.
With these examples, we show clearly how
the field theory observables get
%c4 theirs
their
thermal properties from the bulk despite the
%c3 absent
absence of a black hole,
with the  role of the black hole horizon  played by the cosmological
%c3 horizon,
expansion
%c3
of the de Sitter factor of the AdS metric.

\end{abstract}

\end{titlepage}
\newpage

%c \newpag\tableofcontents

\section{Introduction}

de Sitter spacetime is an important background in cosmology because it
not only describes the late time cosmology, but it is also crucial to
the description of the inflationary early universe. In certain
approximation, one may decouple quantum  gravity and consider the
quantum dynamics of other fields on a background de Sitter spacetime.
Using perturbative quantum field theory on de Sitter space
\cite{birrelldavies}, one
could connect of cosmological perturbation of the CMB in terms of
the quantum fluctuation of the field theory in a slow roll potential.
Among other things,
the prediction of a scale invariant spectrum is in excellent agreement
with the observational results of CMB and  marked a remarkable success
for the inflationary scenario. Nevertheless the picture suffers from
the $\eta$ problem for the inflaton mass. Minimally coupled massless
scalar field also suffers from large secular infrared effects and it
would be nice to have better nonperturbative techniques to deal with
them, beyond the often practiced approximation methods such as
dynamical renormalization group (dRG) \cite{dRG}
or stochastic analysis \cite{sto}.

Recently, by employing a specific dS-slicing coordination of the AdS
space, a duality between type IIB string theory on AdS${}_5
\times$S${}^5$ with dS${}_4$ boundary  and the  $\cN=4$ maximal
superconformal Yang-Mills theory (SCYM) has been proposed \cite{Chu:2016uwi}. %c5
It should be mentioned that while it is not possible to construct global
supersymmetric field theory on four dimensional de Sitter spacetime
\cite{pilch1985,Lukierski:1984it}, the employment of global
superconformal symmetry makes it possible. The Lagrangian of the SCYM theory has been constructed in
\cite{Anous:2014lia}. See also \cite{Hristov:2013spa}-\cite{Klare:2013dka}
for related works. 
The SCYM theory is a cousin of the $\cN=4$ supersymmetric Yang-Mills
theory on flat space and, based on the holographic duality,
it has been argued that SCYM theory would also share certain remarkable
properties like its cousin,
such as exact conformality, $SL(2,Z)$ strong-weak duality
and integrability in some of its sectors. This makes
the studies of the quantum $\cN=4$
SCYM field theory a well motivated problem and it
will be the subject of a different paper.

Here we simply mention that
the de Sitter field theory has a finite temperature $T = H/(2\pi)$ due
to Hawking radiation in de Sitter space. However this temperature has properties
quite different from that of a  thermal field theory in flat
spacetime: 1.
While ordinary temperature break Poincare supersymmetry,
the de Sitter temperature does not break
de Sitter superconformal symmetry. This is partially because the de
Sitter temperature is not a independent parameter but is fixed
directly in terms of the de Sitter space. 2. In terms of holography,
the thermal vacuum of a quantum field theory in flat spacetime is dual
to a black hole deep in the bulk. The presence of a black hole, in particular
its horizon, changes the behavior of bulk supergravity solutions
compared to the case without, and this is how the bulk gravitational
dynamics could account for the properties of the thermal field
theory. However there is no such black hole in
%c3 the
our
dual AdS spacetime.
A motivation of this paper is to understand how the thermal properties
of the de Sitter field theory is encoded in the bulk. It is natural to
suspect that the role of the black hole horizon would be played by the
cosmological
%c3 horizon.
expansion of the AdS bulk. We will demonstrate with examples that
this is basically correct. Nevertheless the way the cosmological
%c3 horizon
expansion affects the bulk solutions are different from the black hole
horizon. In particular, in contrast with the effect of the AdS black
hole where the presence of the gravitational attraction of the black
hole pulls the string along the radial holographic direction, the
cosmological
%c3 horizon
expansion pulls the string in the directions orthogonal to the
radial direction.

As said, the presence of de Sitter temperature is compatible with the
superconformal symmetry of de Sitter space. As the two
and 3-points correlation functions of the theory are completely fixed
by the conformal symmetry, this means they depend on the temperature in a trivial way, through the geodesic distance of
the space. %dg6 completely independent of the temperature.
This is indeed what we
found in
%c2 \cite{chu-ga}
\cite{Chu:2016uwi} using the bulk-to-boundary formalism. However
this is not the case for other more nontrivial observables. For
example, if we consider a Wilson loop operator $W_C$ on the
de Sitter field
theory, the expectation value of the Wilson loop could depend
nontrivially on dimensionless combinations such as $LT$ or $AT^2$,
where $L (A)$ is the length (area) of the loop $C$. This is
a highly nontrivial problem, especially in the strongly coupled regime.
The analysis of such nontrivial
temperature dependence in strongly
coupled field theory in de Sitter space is another motivation of this
paper. %dg6
Notice that gauge/gravity duals in de Sitter where also studied in \cite{Hawking:2000da,Buchel:2002wf,Buchel:2002kj,Aharony:2002cx, Balasubramanian:2002am,Alishahiha:2004md, Ross:2004cb,Balasubramanian:2005bg, Hirayama:2006jn, He:2007ji,Hutasoit:2009xy,Marolf:2010tg, Blackman:2011in, Li:2011bt, Anguelova:2014dza, Zhang:2014cga, Vaganov:2015vpq}
%c3 Another motivation of the paper is to study and analysis the
% thermal properties of strongly coupled quantum field theory on
% de Sitter spacetime using holography.

The plan of the paper is as follows. In section 2, we consider the
computation of entanglement entropy in the de Sitter theory in
arbitrary dimensions. For the 2 dimensional case, an exact analytic
solution is possible.  As a by-product, we use the result of the
entanglement entropy to obtain the equal time
two-point correlator function. The result agrees with the one fixed by
conformal symmetry.
In section 3, we set up a heavy quark bound state with
constant inter-quark distance.
We compute the energy of the system and find that
it behaves similarly to those of a thermal field theory in a flat
spacetime. We discuss in what ways the cosmological
%c3 horizon
expansion and the
black hole horizon differs in
%c3 the
their effects on the bulk dynamics of strings.
In section 4, we consider a spinning heavy quark bound
state with constant inter-quark distance. We regularize the energy and
the angular momenta using the Legendre transformed action. We find
that there exist a maximum spin
%c3 that the state can have in order to be observed.
beyond which the bound state ceases to exist.
This is similar to the behavior of the bound states in a finite
temperature field theories in flat spacetime. Section 5 contains our
conclusions and discussions.

\section{Entanglement Entropy}
%c in $AdS/dS$ slicing}

In this section, we compute the entanglement entropy for
a rectangular stripe on dS space
using holography \cite{Ryu:2006bv,Hubeny:2007xt}.
With a dS-slicing, the metric of AdS space takes the form
\be\lab{metricads}
ds^2_{AdS_{d+1}}=dz^2+\sinh^2 H z ds^2_{dS_{d}}~,\qquad
ds^2_{dS_{d}}=\frac{1}{H^2\eta^2}\prt{-d\eta^2+ dx_i^2}~,
\ee
where $i = 1, \cdots, d-1$, $H=1/R$. In this coordinate patch, the boundary of the AdS space is located at $z =z_{\rm bdy}=\infty$ and is
given by a $d$-dimensional  de Sitter space.
We consider  a rectangular strip $\S$ on the dS space described by
\be\lab{boundary0}
-\frac{L}{2}\leq x_1\leq \frac{L}{2}~,\qquad 0\leq x_j\leq L_1~,
\quad (j =2, \cdots, d-1)~,
%c1 \qquad z\prt{-\frac{L}{2}}=z\prt{\frac{L}{2}}=z_{boundary}~,
\ee
where $L$ is the width of the strip and
$L_1\gg L$ so to ensure translational invariance along the $x_j$
space directions. To obtain the entanglement entropy, we need to
compute the area of the codimension 2 minimal surface $\g_\S$ whose
boundary is given by $\S$.
It turns out sufficient to consider a static
parametrization of the surface
\be\lab{param}
x_1=\s~,\qquad x_j=\s_j~,\qquad  z=z\prt{\s}~,\qquad \eta=\eta\prt{\s}~.
\ee
The boundary condition  at $\s = \pm L/2$ is
\be\lab{boundary1}
z\prt{-\frac{L}{2}}=z\prt{\frac{L}{2}}=z_{\rm bdy}~.
\ee

We remark that the metric \eq{metricads}
belongs to a more general class of bulk metric of the form
\be\lab{metricgen}
ds^2=g_{\eta\eta}\prt{\eta,z}d\eta^2+
%c3
\sum_{i=1}^{d-1}g_{ii}\prt{\eta,z}dx_i^2+g_{\r\r}\prt{\eta,z}
dz^2~,
\ee
where one can set up the computation of the minimal surface in an
uniform manner \footnote{Here the boundary of the
manifold is assumed to be located at $z_{\rm bdy}$ where
$ g_{ii}\prt{\eta,z_{\rm bdy}}=\infty$.}.
The entanglement entropy of the chosen region
\eq{param} is given by
\be\lab{a1}
4 G_N^{(d+1)} S={\rm Area}(\g_\S) =L_1^{d-2}\int d\s A\prt{\s}\sqrt{D}~,
\ee
where
\be\lab{def1}
A:=\sqrt{g_{22}g_{33}\ldots g_{dd}}~,\qquad D:=g_{11}+g_{zz}z'^2+g_{\eta\eta} \eta'^2~.
\ee
The Hamiltonian is a constant of motion, since there is no explicit
$\s$ dependence. Setting it equal to $-c$ we obtain the
constraint
\be\lab{constrain1}
\frac{A g_{11}}{\sqrt{D}}=c~,
\ee
 which gives a first order ordinary differential equation of $z$ and $\eta$
\be\lab{eom1}
g_{zz}z'^2+g_{\eta\eta}\eta'^2+g_{11}\frac{c^2-A^2g_{11}}{c^2}=0~.
\ee
The other two equations of motion are obtained by the variation of the
action \eq{a1} and gives
\be\lab{eom2}
\sqrt{D} \pa_{\a}A +\frac{A}{2\sqrt{D}}\prt{\pa_{\a}g_{11}+
  \pa_{\a}g_{zz} \, z'^2+ \pa_{\a}g_{\eta\eta} \, \eta'^2} -
\begin{cases}
\pa_\s\prt{\frac{A}{\sqrt{D}}g_{zz}z'}=0~,\\
\pa_\s\prt{\frac{A}{\sqrt{D}}g_{\eta\eta}\eta'}=0~,
\end{cases}
\ee
where $\del_\a = \del_\eta$ or $\del_z$.
Eliminating the square root using the constraint
\eq{constrain1}, we obtain
%c1 the simpler version of the equations:
\bea \label{eoms_gen}
\frac{g_{11}}{2 c}\pa_{\a }(A^2)+\frac{c}{2
  g_{11}}\prt{\pa_{\a}g_{11}+
\pa_{\a}g_{zz} \, z'^2+\pa_{\a} g_{\eta\eta} \, \eta'^2}
+\begin{cases}
-c\pa_\s\prt{\frac{g_{zz}}{g_{11}}}z'-c\frac{g_{zz}}{g_{11}}z''=0~.\\
-c\pa_\s\prt{\frac{g_{\eta\eta}}{g_{11}}}\eta'-c\frac{g_{\eta\eta}}{g_{11}}\eta''=0~.\\
\end{cases}
\eea
The desired minimal surface $\g_\S$ is obtained from solving
the differential equations \eq{constrain1} and
\eq{eoms_gen}, subjected to the boundary conditions \eq{boundary1}.

For the AdS${}_{d+1}$ metric \eq{metricads},
the function \eq{def1} and the expressions appear in the equations of
motion are
\be\label{input1}
A(\s)=\prt{\frac{\sinh H z}{H \eta}}^{d-2}~,\qquad
\frac{g_{zz}}{g_{11}}=\frac{H^2 \eta^2}{\sinh^2H
  z}~,\qquad\frac{g_{\eta\eta}}{g_{11}}=-1~.
\ee
Then the Hamiltonian constraint \eq{constrain1} and the equations of
motion \eq{eoms_gen} take the relatively compact form
%c1 please check factors of H %dg6 fixed
\bea
&&z'^2-\frac{\sinh^2H z}{H^2 \eta^2}\eta'^2+\frac{\sinh^2H
  z}{H^2\eta^2}\prt{1-\frac{1}{c^2}\left(
\frac{\sinh^{2}H z}{H^{2}
    \eta^{2}}
\right)^{d-1}
}=0~, \label{eom_z1}
\\
&&z''-2H \coth(Hz) \, z'^2+\frac{2}{\eta}z'\eta'+\frac{\sinh (2 H z)}{2 H
  \eta^2}\eta'^2-\frac{ \sinh(2 H z) }{2H
  \eta^2}\prt{1+\frac{d-2}{c^2}
\left(\frac{\sinh^{2}H z}{H^2 \eta^2}\right)^{d-1}
} =0~, \nn \\ \lab{eoms_ds} \\
&&\eta''+\frac{1}{\eta}\eta'^2-\frac{1}{ \eta}\prt{
1+\frac{d-2}{c^2}\left(\frac{\sinh^{2}H z}{H^2 \eta^2}\right)^{d-1}
}=0~. \label{eom_eta}
\eea %dg6 fixed
In arbitrary number of dimensions the equations can be solved
numerically. However, for the case of 2 dimensional conformal field
theory, the factor $A\prt{\s}$ in
\eq{input1} is equal to the unit and the equations of motion can be
solved analytically.

\subsection{Entanglement entropy for AdS${}_3$/dS${}_2$}

Let us consider the case of a two dimensional conformal field theory.
The equation of motion  \eq{eom_eta} for $\eta$ reads
\be\lab{eom2d1}
\eta''+\frac{\eta'^2}{\eta}-\frac{1}{\eta}=0~.
\ee
This has the solution
\be\lab{sol2d1}
\eta=\pm \sqrt{\s^2+c_1\s+c_2}~,
\ee
with $c_{1,2}$ being  constants of integration. Notice that the
solution of $\eta$ does not depend on the cosmological horizon
$H$. Similarly the equation \eq{eoms_ds} gives
\be%dg6 fixed
z''- 2H \coth (H z) z'^2+\frac{2}{\eta}z'\eta'
-\sinh(2Hz) \frac{1-H^2 \eta'^2}{2 H \eta^2}=0~
\ee
and by substituting the solution \eq{sol2d1},
we get an second order differential equation for $z\prt{\s}$
\be\lab{eom2d2}
z''-2H \coth (H z)\, z'^2 + z' \frac{2 \s+c_1}{\s^2+c_1\s+c_2}+\sinh
(2 H z)\frac{c_1^2-4 c_2}{8 H \prt{\s^2+c_1\s+c_2}^2}=0~.
\ee
The solution for $z$ then can be written in the compact form as
\be\lab{sol2d2}
%c1 factor of H
\coth H z=\pm\frac{c_3}{4} \cdot \frac{4+c_4\prt{c_1^2-4c_2} \s}
{\sqrt{c_2+c_1\s+\s^2}}~,
\ee
where $c_3$ and $c_4$ are arbitrary integration constant.
To specify the integration constants, we note that due to the
symmetry of the problem, the geodesic should be left-right symmetric
with respect to $x_1$.
The geodesic has a turning
point in the bulk $(z_0,\eta_0)$, which by the symmetry of the space,
must therefore be located at $\s=0$, and the
%c3
desired functions $z$ and $\eta$
must be even functions of $\s$. We obtain immediately that
\be\lab{cc1}
\eta_0=\pm \sqrt{c_2}~,\quad c_1=0~, \quad c_4=0
\ee
and
%c1
\be \label{etaa}
\eta = \sqrt{\s^2 + \eta_0^2}~.
\ee
In addition, $c_3$ can be expressed in terms of the turning point
coordinates as
\be \lab{cc2}
\coth^2 H z_0=\frac{c_3^2}{\eta_0^2}~.
\ee
Now the boundary condition \eq{boundary1} at $ \s = \pm \frac{L}{2}$ gives
\be\lab{cc2b}
c_3^2 =\frac{L^2}{4}+\eta_0^2~.
\ee
Therefore, eliminating $c_3$,
we get the relation between the turning point  $(z_0,\eta_0)$ and $L$:
%c1 length of the strip $L$:
\be\lab{initial_data}
\frac{\sinh^2 Hz_0}{\eta^2_0}=\frac{4}{L^2}~.
\ee
%c1 The bulk data of the saddle points of the surface is combined in a
% conformal factor of the metric \eq{initial_data} to specify the length
% $L$ of the strip and therefore the entanglement entropy. This is
% because our boundary is conformally flat. This also implies that
% strips of length $L$, for different boundary times $\eta(L/2)$ can
% have the same entanglement entropy.
Note that
the Hamiltonian constraint \eq{constrain1} is satisfied by the
solutions and applying it at the turning point for the geodesic it
fixes the constant c as
\be\lab{cc3}
c^2=\frac{4}{H^2 L^2}~.
\ee
The minimized action is
%c1 after the integration takes the form
\be \lab{ee2d}
4 G_N^{(3)} S=\frac{2}{H} ~\mbox{arctanh} \frac{2
  \s}{L}\bigg|_0^{L/2-\ve}=
%c2
R \log\prt{\frac{L}{\ve}}~,
%c1 R \log \frac{L}{\ve}~,
\ee
where $R:=1/H$ is the radius of the AdS space  and
$\ve  \approx 0$ is %c2
a UV cutoff
imposed in the $\s$ direction of the worldsheet.
%c2 on the boundary field theory.
%c2
To compare with the field theory, we need to express \eq{ee2d}
in terms of the UV cutoff of the field theory. This can be achieved by
noting that if we introduce the
%c3
radial
coordinate $\rho$ of the bulk defined by %dg3
\be\label{zrho1}
dz = R \frac{d\rho}{\rho}~,
\ee
then the desired UV cutoff is given by
\be \lab{eps2}
R \log \frac{1}{\e} :=z\prt{{\frac{L}{2}-\ve}}\simeq\frac{1}{2 H}\log
\frac{4 \eta_b^2}{L \ve}~,
\ee
where
\be \label{etab}
\eta_b := \eta \prt{\frac{L}{2}} =
  \sqrt{\frac{L^2}{4}+\eta_0^2}~,
\ee
is the time coordinate for the boundary point of the string.
As a result,
\be \label{S-etab}
4 G_N^{(3)} S = 2 R \log \frac{L}{\e \eta_b}~.
\ee
As $\eta_b$ should be treated as an independent parameter apart from
$L$,  we obtain \cite{Maldacena:2012xp,Fischler:2013fba}
\be
\lab{entdsf}
S= \frac{c}{3} \log \frac{L}{\e}~, \qquad c=\frac{3 R}{2G_N^{(3)}}~.
\ee
%c3
The  entropy \eq{entdsf} has a trivial $L$ dependence as
fixed by conformal
symmetry and has a
trivial dependence on the de Sitter temperature.
We also note that \eq{entdsf}
is the same as the result in the flat space. This must be the case
since the boundary metric is conformally related to that of the flat
space and the logarithmic divergent piece is universal 
and depends trivially on the de Sitter temperature through the geodesic distance.
%dg6 and independent of this conformal factor.

\subsection{Two-Point Correlation Function}

In the above computation of the holographic entanglement entropy, we
have computed the length
$L_g$ of the bulk geodesic joining the two points
$x_1= \pm L/2$ on the boundary.
As an application, this can be used to reconstruct the boundary
correlation function for conformal operators. In general, for 
a scalar field of mass $m$ in the bulk of AdS${}_3$, it has the lowest
energy eigenvalue $H \D$ and
the bulk propagator from $x$ to $x'$ is given by
\be \label{Gbulk}
G(x,x') =\int\cD\cP e^{- H \D L\prt{\cP}},
\ee
where $\cP$ is a path joining the two points and $L(\cP)$ is the
proper length of the path $\cP$. In the semiclassical limit, the path
integral is localized to its saddle points and is given by a
sum over the geodesics. In the present case
\be \label{Gbulk1}
G(x,x') = e^{- H{\D} L_g}~.
\ee

According to
\cite{Balasubramanian:1999zv,Banks:1998dd}, \eq{Gbulk} is  also equal
to the
CFT correlator for the dual operator $\cO$ in the large $N$ limit.
Therefore we obtain in the large $N$ limit and semiclassical
approximation the following expression for the two point function
\be
\vevv{\cO\prt{x}\cO\prt{x'}} \simeq
%c2 e^{\frac{2\D H}
\prt{\frac{1}{\e}}^{2 \D } e^{-H \D  L_g}~,
\ee
where we have regulated the two point function by
%c3 multiplying it
adopting a normalization involving
an appropriate expression of the cutoff $\e$.
%c3 This can be considered as a particular choice of the normalization
%of the two point function.
%c2 Note that the conformal time $\eta$ for the boundary points is given by
% \be \eta_b:= \eta\prt{\pm \frac{L}{2}} =
% \sqrt{\frac{L^2}{4}+\eta_0^2}~. \ee
Using \eq{S-etab}, we obtain
\be\lab{vevfinal}
\vevv{\cO\prt{-\frac{L}{2},t}\cO\prt{\frac{L}{2},t}}
%c1 \simeq
=
\prt{\frac{\eta_b}{L}}^{2\D} \sim \frac{1}{\s^{2\D}}~,
\ee
where $\s$ is the de Sitter invariant distance
\be\label{invariant}
\s^2\prt{x,x'}:= \frac{-\prt{\eta-\eta'}^2+\prt{x-x'}^2}{H^2 \eta \eta'}~.
\ee
\eq{vevfinal} agrees with the result obtained in \cite{Chu:2016uwi}
using the bulk-to-boundary formalism.
As the result is completely determined by conformal invariance, %dg6 independent of the de Sitter temperature
it is trivially depending
on the de Sitter temperature through the geodesic distance.

%c1 \section{The space-like Wilson loop in AdS/dS Gauge/Gravity duality}

\section{Static Mesons in dS Theory}

To see nontrivial dependence on temperature,
let us introduce heavily massive external quarks and consider
expectation value of the Wilson loop operators for space-like loop $C$
in the dS conformal field theory.
According to holography \cite{adscft1,adscft2} %dg6 I put the correct refs
%c2 \cite{malda},
they are determined in the large $N$ limit,
by  the minimal surface formed by the string world-sheet ending on the
loop $C$ on the dS boundary.
%c1 \subsection{The $AdS/dS$ Wilson Loop from Holography}
For convenience let us go to the
planar coordinates for the dS space by setting $\eta=e^{-2 H t}$, we have
\be\lab{metricads2}
ds^2_{AdS_{d+1}}=dz^2+\sinh^2 H z~ ds^2_{dS_{d}}~,\qquad
ds^2_{dS_{d}}=
-dt^2+ e^{2 H t} dx_i^2 ~.
\ee

%c1
\subsection{The String Solution}

We consider  quark, anti-quark pair at the boundary \eq{metricads2} at
\be\label{bcwl}
t=\t~,\qquad x_1= \pm \frac{L}{2} e^{-H t} ~,
\ee
where the $\pm$ sign correspond to the positions of Q and \={Q}
respectively.
Note that  in contrary to the flat space case, we have specified a
specific time dependence for the position of the quarks, which give them
a constant speed
\be\label{velocity}
v=\pm\frac{H L}{2}~.
\ee
pointing towards each other. This counter balance the expansion of
the dS space and results   in a
constant $dS$ invariant distance \eq{invariant}
between the quarks
\be
%c3 \rm
\s_{\rm inv}^2(Q,\bar{Q})= L^2~.
\ee
In other words, we have chosen here to consider meson of constant
size and this is the closest analogy to the flat space case.
Motivated by \eq{bcwl}, we parametrize the string worldsheet as
\be\label{wlansatz}
t=\t~,\qquad x_1=  e^{-H t} \s~,\qquad z=z(\s)~.
\ee
%c3 allowing to the string profile in the bulk to be determined by the
% equations of motion.
As we will see below, this parametrization
guarantees  time translation invariance for the Wilson loop and
greatly simplifies
the problem since the string world sheet is then govern by  ordinary
differential
equations instead of partial differential
equations.
This is not the case if we have considered
the usual static gauge parametrization $t=\t,~x_1=\s,~z=z(\s)$, %dg6
which does
not satisfies the equations of motion.

%c1 \subsection{The Properties of the Bound State}
%c1 \subsubsection{The String Solutions}
It is not difficult to check that the Nambu-Goto
%c3
(NG) action
\be
S=\frac{\sqrt{\l}}{4\pi} \int d\s d\t \sqrt{-g}~,
\ee
for the parametrization \eq{wlansatz}
%c3
is consistent and
gives only one non-trivial equation of motion
\bea
&&(1-H^2\s^2) \sinh^2 (H z(\s))~ z''(\s)-  H^2 \s (1-H^2\s^2) ~ z'(\s)^3
\nn\\
&&
-\frac{3H }{2} (1-H^2\s^2)\sinh (2 H z(\s)) ~z'(\s)^2
-2 H^2 \s\sinh^2 (H  z(\s))~ z'(\s)\nn \\
\label{eomwl}
&& -2 H \cosh (H  z(\s)) \sinh^3 (H z(\s))=0~.
\eea
The on-shell action takes the compact form
\be\lab{onshellaction1}
\frac{4\pi}{\cT \sqrt{\l}} S=\int d\s \sinh H z\sqrt{\sinh^2 H
  z+(1-H^2\s^2) z'(\s)^2} ~,
\ee
where we have integrated the world-sheet time to give a factor of
$\cT$.
Notice that the resulting action is time independent  reflecting the
fact that the endpoints of the string have static invariant
distance.
%c1 The Legendre transform adds only a boundary term so the
% equation \eq{eomwl} is not modified by its presence.
For $H \s>1$ the action \eq{onshellaction1} may become zero or
imaginary, this is a common characteristic for the orthogonal Wilson
loop action in finite temperature field theories, where $H$
%c3 there
here plays the role of the black hole horizon.

\begin{figure*}[!ht]
\begin{minipage}[ht]{0.5\textwidth}
\begin{flushleft}
\centerline{\includegraphics[width=75mm]{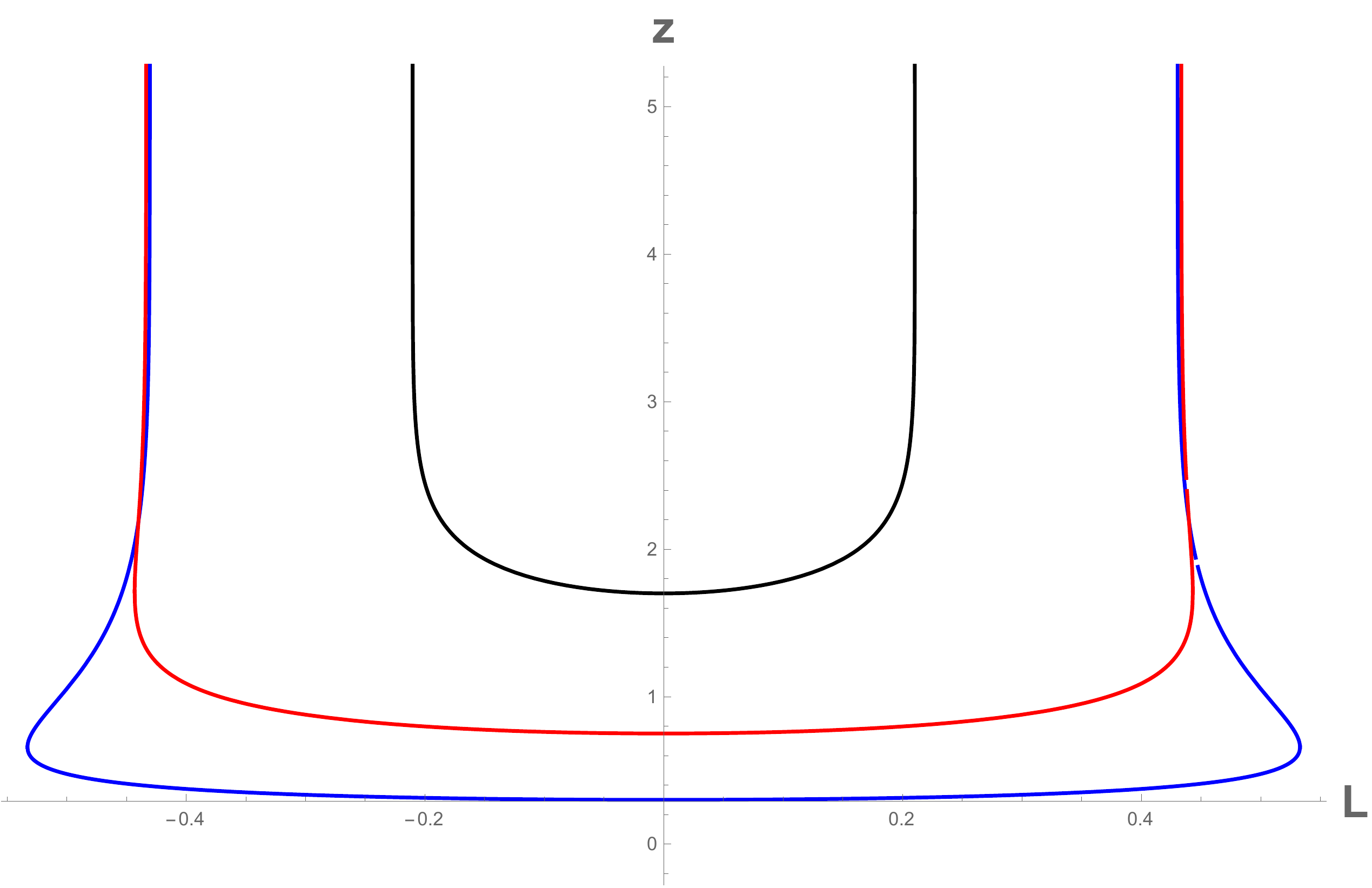}}
\caption{\small{The string profile for different values of the
    inter-quark distance
 $L$. The boundary is at the large values of $z$. We plot string
    solutions with three different turning points $z_0$. The one with
    smallest $L$ has a string profile that is minimally deformed by
    the cosmological expansion. %horizon. 
     The other two string solutions
    correspond to the same inter-quark distance $L$, while the
    acceptable one has an endpoint closer to the boundary and is the
    energetically favorable. The string solution that goes deeper to
    the bulk, is clearly deformed by the cosmological expansion. %dg6 horizon.
    }}
\label{figure:a1}
\end{flushleft}
\end{minipage}
\hspace{0.3cm}
\begin{minipage}[ht]{0.5\textwidth}
\begin{flushleft}
\centerline{\includegraphics[width=75mm ]{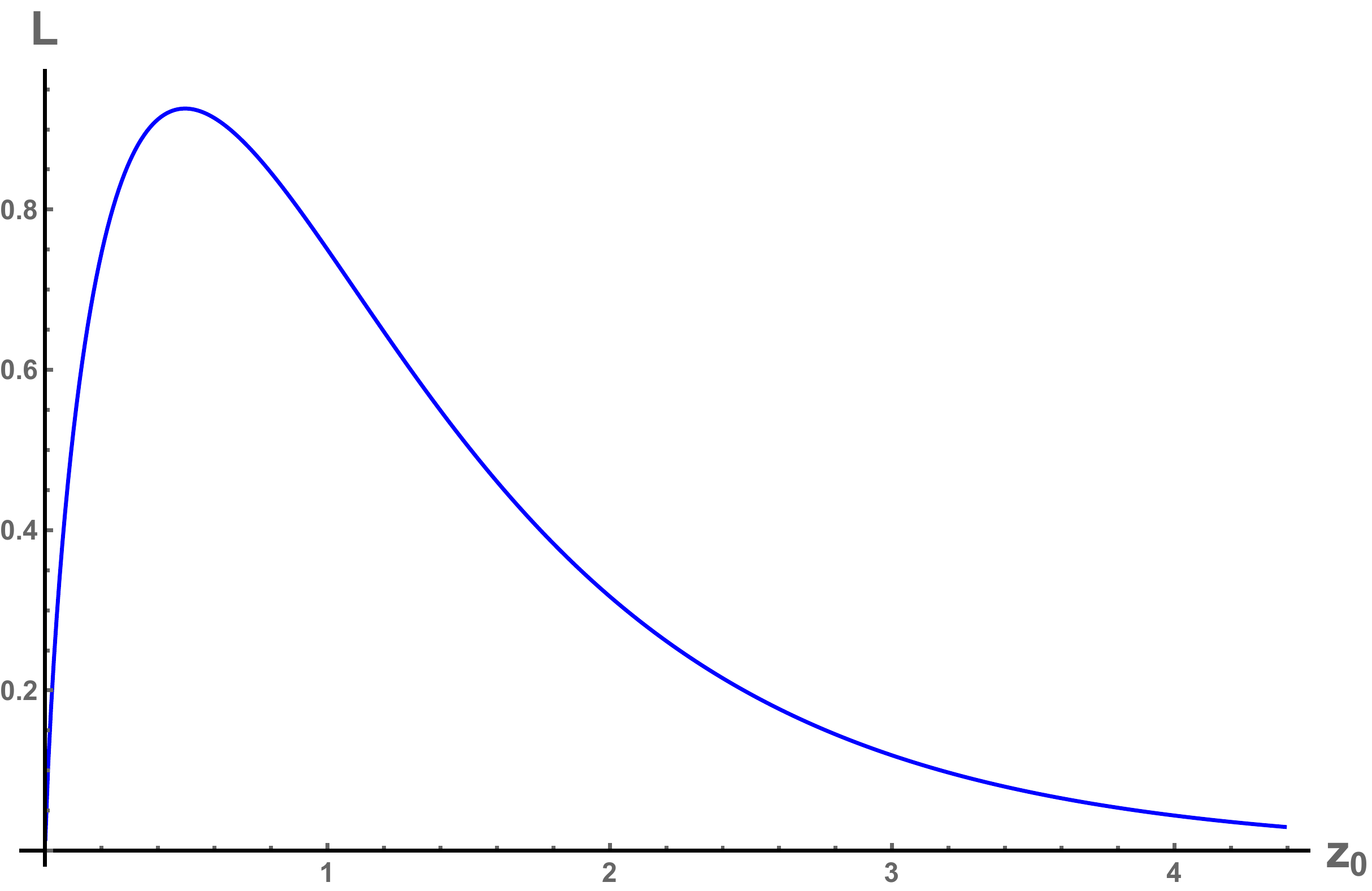}}
\caption{\small{The inter-quark distance $L$ in units of $H$, in terms
    of the turning point $z_0$ of the string. For each value of $L$
    there exist two string solutions with different turning points and
    energy. The acceptable one lies on the right branch of the maximum
    of the curve, since it has lower energy and it is stable. An
    example of such twin-solutions with the same boundary conditions
    is presented in Figure \ref{figure:a1}.} }
\label{figure:a2}\vspace{1.4cm}
\end{flushleft}
\end{minipage}
\end{figure*}

The solution of the equation \eq{eomwl}
%c1 determines the string profile in the bulk
can be found numerically and is presented in Figure\footnote{All
% the quantities in
plots are in units of $H=1$
%c1 , i.e. $L H$. All our plots
%  are for $H=1$,
unless otherwise stated.} \eqref{figure:a1}. For
small inter-quark distances $LH<<1$ the profile of the string has the
usual U-shape
%c1 , similar to
of a hanging chain form with two fixed end-points.
%c1 It is similar the world-sheet profile in the static
% backgrounds.
As the distance between the pair increases,
the gravitational effects of the cosmological
%c3 horizon
expansion in the interior of the AdS space
give rises  to a deformed U-shaped profile
with more substantial modification
around its turning point.

A couple of remarks are in order:

1. A common property of the holographic finite temperature field theories
is that to each boundary distance of the string, there are two string
solutions extending inside the bulk with different radial
%c1 distances
dependence
\cite{Rey:1998bq,Brandhuber:1998bs}.  This is
also  the case here.
%c1 of the string world-sheet in gravity dual theory of a $dS$ field
%theory.
As can be seen in  Figure \ref{figure:a1},
there are two different string profiles in the bulk with
different turning points $z_0$ for each boundary separation $L$.
In Figure \ref{figure:a2}, we plot $L$ as a function of the turning
point $z_0$. We find that for inter-quark distance less than a certain
maximum, namely
\be \lab{lmaximum}
%c3 \rm
L_{\rm max} H \simeq 0.92~,
\ee
there always exists two different
connected string solutions.
%This is a indication that the string feel the present of the
%cosmological horizon at the boundary.
%c1
\footnote{A similar observation to \eq{lmaximum} has been made in
\cite{Fischler:2014ama}, where we comment on among other discussion in
the Appendix A.} 
%c1
At this point a natural question arises for the reason of the
resemblance of our string solutions
%c1 corresponding to the $dS$ heavy quark probes,
with the ones in the
%c1
usual holographic
thermal field theories. The
mechanism that provides here twin world-sheets solutions
is not
because of the presence of a black hole in the bulk
%c3
but is due to the expansion of the AdS space, an
effect which is visible when the AdS metric is expressed in the
form \eq{metricads} in terms
of the dS-sliced coordinates.
%c3
Effectively, we have placed the string in the AdS space
while keeping fixed the distance of the two boundary string endpoints by
counterbalancing the expansion of the space  with a given boundary
velocity. However the rest of the string
%c3 dynamics are
is still affected  by the expansion, where the effect
is enhanced
%c3 away from the boundary endpoints.
as one goes deeper in the bulk.
At some point the
deformation of the connected string becomes so large that the string
prefers energetically
%c3, to break to
to break and become two separate strings and this is
when the heavy quark bound state
%c3 stops to exist.
dissociates.

%c1 So instead of having only the zero temperature U
% shaped string world-sheet solutions, we obtain another twin solution
% that goes deep to the bulk, it is deformed by
% the effective potential and has larger energy. %This is a way to
                                %partly justify the thermal behavior
                                %of our string world-sheet.

\begin{figure*}[!ht]
\begin{minipage}[ht]{0.5\textwidth}
\begin{flushleft}
\centerline{\includegraphics[width=75mm]{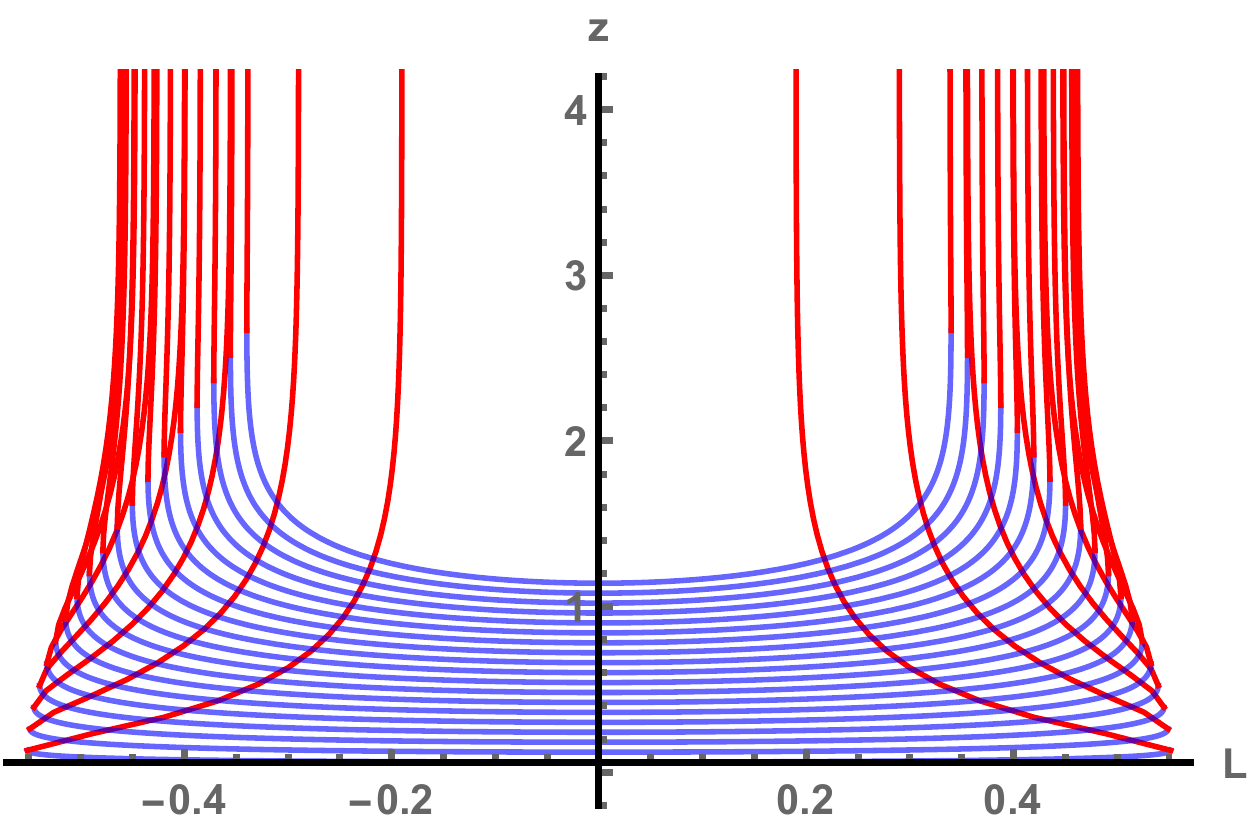}}
\caption{\small{The string worldsheet for different values of the
    inter-quark distance $L$. We observe the changes on the solution
    as the turning point $z_0$ moves away from the boundary. In the
    bottom blue-colored region the derivatives $z'\prt{\s}$ are
    finite, while the red-colored part includes the point of the
    infinite $z'\prt{\s}$ derivative. Notice the string deformation
    enhancement as it goes closer to the bulk.}}
\label{figure:a3}
\end{flushleft}
\end{minipage}
\hspace{0.3cm}
\begin{minipage}[ht]{0.5\textwidth}
\begin{flushleft}
\centerline{\includegraphics[width=73mm]{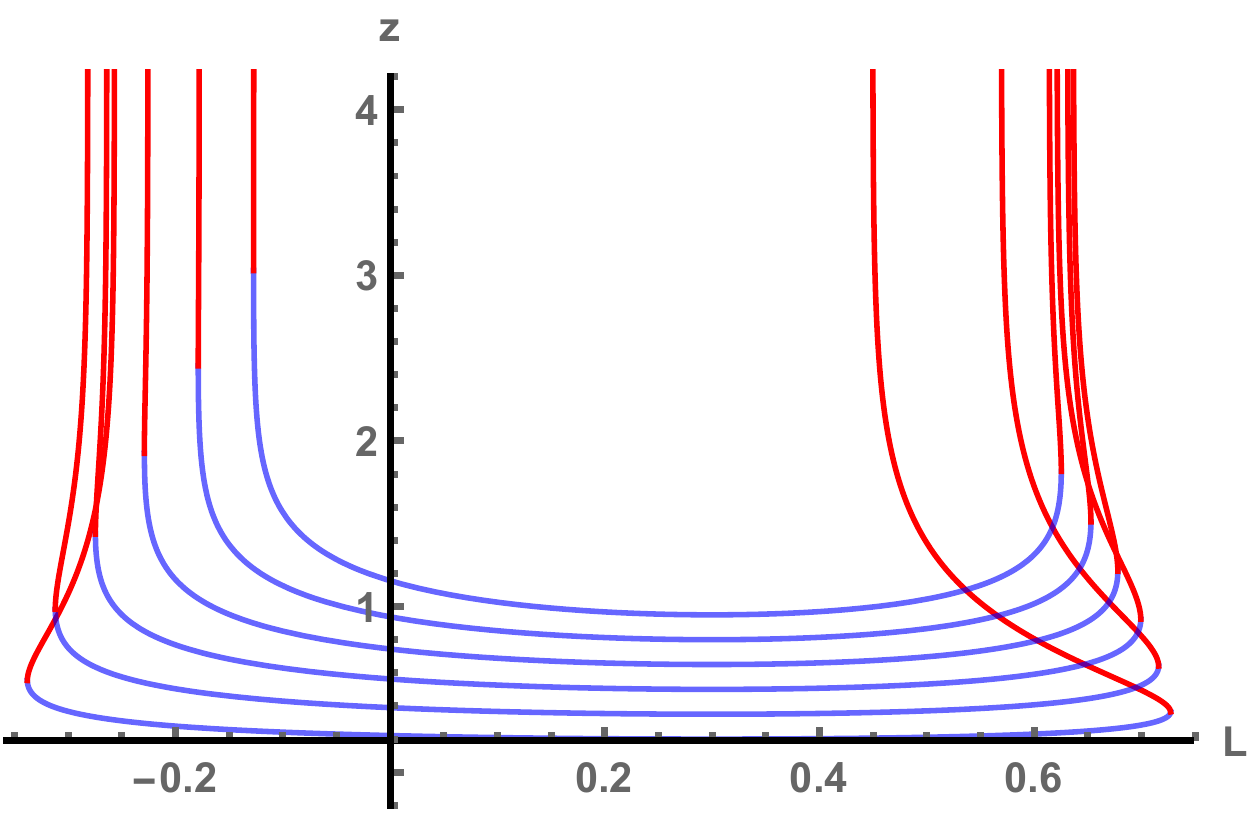}}
\caption{\small{The heavy quark pair is located asymmetrically with
    respect to the origin. This asymmetry generates an asymmetric
    deformation
of the bulk string, enhanced on the side that is further away from the
origin, in this case at the right side. The midpoint of the string is
at $\s_0 H=0.3$.}}\label{figure:a4} \vspace{0.7cm}
\end{flushleft}
\end{minipage}
\end{figure*}

2. The deformation of the string in the bulk due to the
%c3 cosmological horizon
cosmological expansion leads to a minor complication in the
numerics \footnote{Similar complications have been observed to strings
  with rotating endpoints \cite{Peeters:2006iu}. To simplify the
  numeric procedure, one may choose a different gauge in the string
  parametrization.}. To obtain the string profile we solve the
equation \eq{eomwl} by selecting the initial value of the holographic
distance of the turning point of the world-sheet in the bulk $z_0$ and
we shoot from this point towards the boundary. The selection of $z_0$
specifies the inter-quark distance $L$ at the boundary. For $\s=\s_1$
the string has a second turning point at $z_1=z(\s_1)$, where
$z'(\s_1)=\infty$ (Figure \ref{figure:a1}). At this point we need to
invert the differential equation \eq{eomwl} to obtain the equation for
$\s(z)$ and to shoot from the point $\s_1$ with initial conditions
$\s(z_1)=\s_1$ and $\s'(z_1)=0$. The two solutions can be combined to
get the full profile of the string. Notice that we need the full
solutions in order to find the energy of the string, where we
integrate the energy density from $\s=\s_0$ to $\s=\s_1$ as a function
of $\s$, and from $z_1$ to the boundary as a function of $z$. A series
of such solutions are shown in Figure \ref{figure:a3}.

3. Notice that the deformation on the world-sheet due to the cosmological
%c3 horizon
expansion is symmetric in the two edges. This is because we place the Q
and \={Q} at equal distances from the origin $x_1=0$. Non-symmetric
displacement of the pair along the
%c3
$x_1$-axis leads to asymmetric
deformation of the string world-sheet, enhanced
%c3 to
on the side that is
further away from the origin. A representative solution is shown in
Figure \ref{figure:a4}.

\subsection{The Energy of the Bound State}

To obtain the energy of the bound state, one needs to regulate the
on-shell Nambu-Goto
%c3 (NG)
action
which is infinite due to the infinite length of the string
world-sheet. Two subtraction schemes,
%c3
the Legendre subtraction scheme and
the mass subtraction scheme,
%c3 and the Legendre subtraction scheme,
have been widely used.

As the NG action is a functional of
coordinates and the holographic direction of the
string world-sheet satisfies a Neumann
boundary condition, one needs to perform a Legendre transformation
to change the boundary
condition for the modified action. %dg3
We point out that the Legendre transform is not diffeomorphic
invariant;
%c3 add;
and for a successful canceling
of the UV divergences, we should use the coordinate system $\r$
\eq{zrho1} which is
%c3 analog
analogous to the AdS Poincare coordinates. The
Legendre transform of the action in the $\rho$ coordinates is
\cite{Drukker:1999zq,Chu:2008xg} %dg6 added undefined ref
\be\lab{Legendre0}
%c3
S_{\rm Legendre}=\int d\t \r p_\r  \Big|^{\s\rightarrow \s_2}_{\s\rightarrow \s_1}~,
\ee
where $\r$ is the holographic direction, $p_\r$ is the conjugate momentum
\be
p_\r=\frac{\delta S}{\delta (\partial_\s \r)}~,
\ee
and $\s_1$ and $\s_2$ are the endpoints of the string.
Using the equation \eq{zrho1},
%c3
% the conjugate momentum in $z$
% coordinates \eq{metricads2} becomes $\r p_\r\rightarrow p_z$
we obtain $\r p_\r = p_z$, where
\be
p_z = \frac{\delta S}{\delta (\partial_\s z)}~
\ee
is the conjugate momentum in the $z$ coordinate, and
%c3 the modified action that cancels successfully the divergences
the desired Legendre term takes the form
%The modified Lagrangian in Poincare coordinates is
%\cite{Drukker:1999zq}
\be\lab{Legendre}
\tilde{S}= S - \int d\t p_z  \Big|^{\s\rightarrow \s_2}_{\s\rightarrow \s_1}~.
\ee
%where $z$ is the holographic direction,
%$p_z$ is the conjugate momentum
%\be
%p_z=\frac{\delta S}{\delta (\partial_\s z)}~,
%\ee
%and $\s_1$ and $\s_2$ are the endpoints of the string.

%c3
The idea of the mass subtraction scheme is very intuitive.
For the
standard case of $\cN=4$ SYM, a
single heavy quark is the endpoint of a straight string that
initiates from the
boundary of the space ($z= \infty$) and goes into the bulk. The infinite mass of
the quark  is given by
\be\label{adsmass}
M_Q=\int_{0}^{\infty} d z ~.
\ee
The subtraction of this mass from the energy is equivalent to the
subtraction using
the Legendre term. For the  finite temperature $\cN=4$ SYM theory,
the gravity
dual has a black hole which introduces a lower bound on the holographic
coordinate with a range $[z_h, \infty)$, where $z_h$ is the position of
the black hole.  Then the corresponding mass for the heavy quark
is given by
\be\label{adsbhmass}
M_Q=\int_{z_h}^{\infty} d z~.
\ee
Note that apart from allowing to cancel  the UV divergence of the NG
action of the connected string worldsheet, $M_Q$ now also makes
%c3 a
an IR
contribution $z_h$, which is interpreted as a
thermal contribution to the bound state energy.
%c3
On the other hand in
the Legendre term
%c3 \eq{adsmass},
\eq{Legendre},
information about the thermal
properties of the horizon is present
%c3
through the conjugate momentum and the
string solution itself, however in a less direct manner.
When
computed at the boundary, the black hole contribution is negligible and
the Legendre boundary term is equal to that of the zero temperature
theory. In general, while both schemes offer the cancellation of the
UV divergences, they could differ in their finite IR contribution and
two schemes are not equivalent.
The choice of regularization scheme depends on the problem and
the physical quantity one desire to compute.

In the present case, due to its intuitive picture,
one may want to use the mass subtraction scheme by
subtracting out  the energy
of two single non-interacting quarks moving with the velocity
\eq{velocity}. To do this, one needs an appropriate string
solution with a single moving endpoint at the dS boundary
\be\label{staticbc}
x_0=  \t~, \qquad x_1= \frac{L}{2} e^{-H t}~
\ee
with constant velocity. This is however not straightforward to find.
In fact the most straightforward string world-sheet  parametrization
\be
x_0= \t~, \qquad x_1=  \frac{L}{2} e^{-H t}~,\qquad z= z(\s) ~
\ee
is not  a solution of the
%c1 system to the full equations of
full system of equations of
motion, and a more involved string profile required. On the other
hand, the Legendre subtraction can be performed without any problem.
%c1
As a result, the regularized energy for our meson system is given by
\be\lab{Etotalstatic}
E_{\rm tot}(L)= S_{Q\bar{Q}}-2 S_{UV}~,
\ee
where $S_{QQ}$ is given by
%c1 twice
the onshell action \eq{onshellaction1} and $S_{UV}$ is
given by %dg3
\be\lab{legendreapplied}
%c2 please check
\frac{4\pi}{\cT \sqrt{\l}} S_{UV}=  \sqrt{1-\frac{H^2  L^2}{4}}
 \sinh Hz\Big|_{z\rightarrow \infty}~,
\ee
where we have used the fact that our solution is independent of $\t$
to integrate through the time to get an overall factor of
$\cT$. The factor of 2 accounts for the contribution from both
endpoints.
%c3 Note that in obtaining \eq{legendreapplied}, we have used the fact
%c2
We have also used the fact that
near the endpoints $\s= \pm L/2$, $z\to \infty$,
the differential equation \eq{eomwl} is solved by
\be
z' = \mp \frac{ 1}{2HL} e^{2Hz},
\ee
where the sign $\mp$ is for $\s =\pm L/2$.
%dg3 $-(+)$ is for $\s \sim L/2 (-L/2)$.
As a result, near the boundary, $p_z$ is given by
%c3 can be approximated by
\be
p_z = \pm \sqrt{1-\frac{H^2 L^2}{4}} \sinh Hz,
\ee
where the sign $\pm$ is for $\s =\pm L/2$.
%dg3$+(-)$ is for $\s \sim L/2 (-L/2)$
%c2
We remark that
unlike the  AdS case in the Poincare coordinates where
the UV divergences of
the Wilson loop do not depend on the spatial position of the
string, in the present dS-sliced description of the AdS space where
the metric becomes time dependent,
the UV boundary term is multiplied by a factor that depends on
the spatial position of the string. This is essential to cancel out the
infinity.

\begin{figure*}[!ht]
\centerline{\includegraphics[width=73mm]{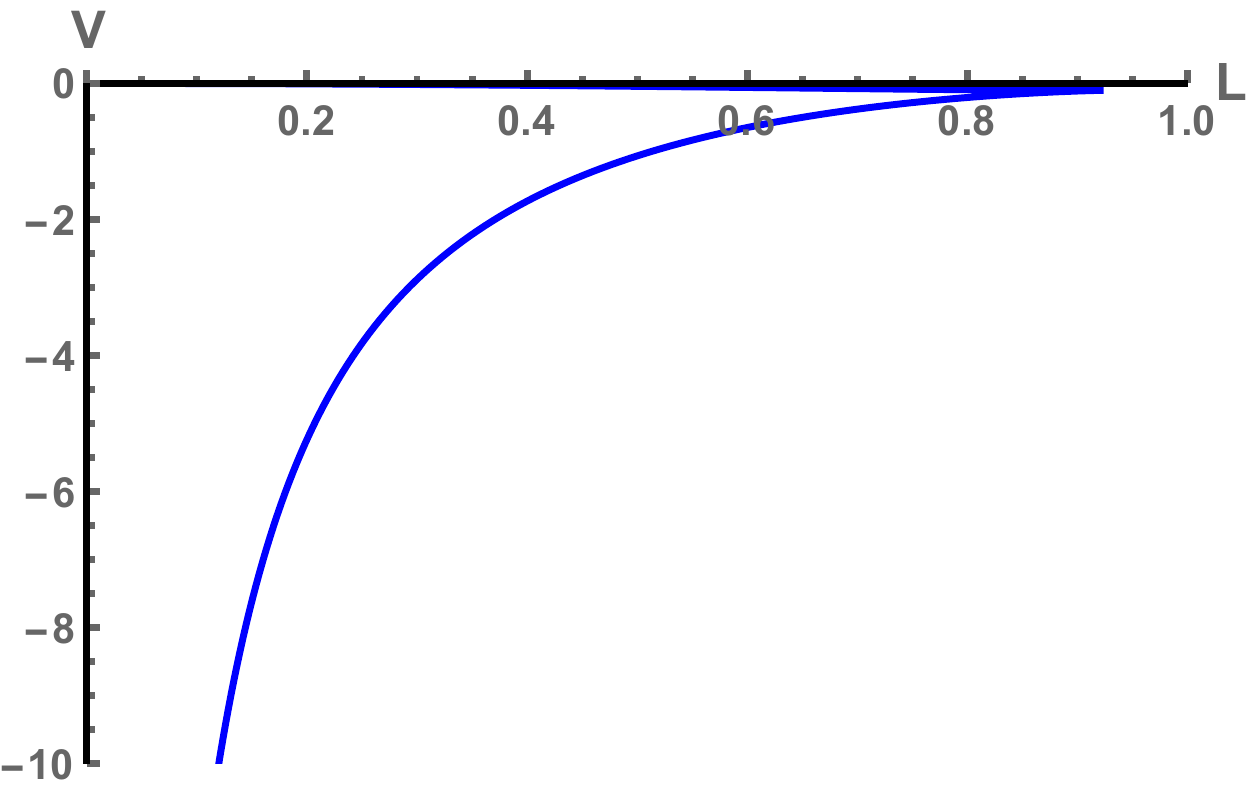}}
\caption{\small{The regularized energy of the bound state in terms of
    the size of the bound state $L$ in units of $H$. We notice that
    there is a maximum value of distance $L$ with a maximal energy,
    beyond which there is no minimal surface with the boundary
    conditions \eq{bcwl}. The turning point occurs for negative values
    of energy, and the almost flat branch corresponds to the
    non-stable solutions that are energetically non-favorable.
}}\label{figure:a5}
\end{figure*}

Our result for $E_{\rm tot} (L)$ is plotted
in Figure \ref{figure:a5}. The regularized energy of the bound state
in the AdS/dS space has similarities with that of the bound state in
the AdS black hole and the dual finite temperature $\cN=4$ SYM field
theory.
The energy $E(L)$ has a turning point, indicating a maximum size of
the heavy quark bound state with maximal energy for the state, beyond
which it does not exist. Moreover, there exist two string solutions
corresponding to the same size meson but have different energy. The
acceptable solution is the one with the minimum energy which
corresponds to the stable and energy preferred state. This resembles
the known holographic results of finite temperature %dg6 filed
field theories
including a black hole horizon. %
%\cite{Brandhuber:1998bs,Rey:1998bq}.
%c4
Notice, however, that the energy of our solution
%c4 is below the $E_{\rm tot}(L)=0$
does not cross the horizontal axis.
This is  different from the behavior in thermal field theory
with black hole dual 
\cite{Rey:1998bq,Brandhuber:1998bs}.
There the mass subtraction scheme was adopted, and it was found that
the energy becomes positive at a certain inter-quarks
separation $L_* < L_{\rm max}$. This signifies a phase transition where
having a pair of straight line strings ending
directly on the horizon of the black hole has become
the energetically more favorable string configuration.
As we do not have a black hole and we have used the Legendre transform scheme
for the cancelation of the UV divergences, %dg6 these kind of
%straight line string solution,
this explains the absence of such phase transition in our case.

%c4 The difference is because
% we are using a different subtraction scheme here. In the mass
% subtraction scheme as adopted in the above mentioned papers,
% the string length is integrated from the boundary to the black hole
% horizon, as in \eq{adsbhmass}, and so the energy receives
% an explicit IR contribution which depends on the
% temperature of the heat bath.
% On the other hand, in the Legendre subtraction scheme \eq{Legendre},
% the endpoints of the string at the endpoint are
% hardly affected by the cosmlological
% expansion and so  thermal mass contribution is not included.

To summarize, we find that
%c3
similar to the situation of mesons in the usual thermal field theory
whose
temperature has a holographic origin in terms of a black hole,
here our mesons also admit a pair of string solutions for each
admissible boundary condition which leads to bound state.
However the responsible mechanism here is different: it is
due to the presence of
cosmological expansion in longitudinal directions parallel to the
boundary, rather than attraction due to the black hole in the
radial/transverse direction \cite{Brandhuber:1998bs,Rey:1998bq}. To elaborate further
on the properties of bound state we add one more degree of freedom to
our system in the next section.

\section{Spinning Mesons in dS Theory}

In this section we examine the spinning mesons modeled by rotating
hanging strings from the dS boundary.

\subsection{The Holographic Setup}

The spinning string we consider, has its two endpoints on the
boundary, corresponding to the quarks of the spinning meson. To
consider rotation, it is convenient to rewrite
the planar coordinates metric \eq{metricads2} in the following
spherical  form
\be\label{spherical}
ds^2=dz^2+\sinh^2 H z~\prt{ -dt^2+e^{2 H t} \prt{d\rho^2+ \rho^2
    d\th^2+  \rho^2\sin^2 \th d\phi^2}}~.
\ee
Without loss of generality, we consider rotation
along the equator of the sphere with
the following parametrization ($0 \leq \s \leq L/2$)
\be\lab{spinning}%dg5
t= \t~,\qquad \th=\frac{\pi}{2}~, \qquad \phi\prt{\s,t}=c+\omega
t-\omega f(\s)~,\qquad \rho\prt{\s,t}=\s e^{-H t}~,\qquad
z=z\prt{\s}~,
\ee
where $c$ is a constant to be determined by the two points of the boundary.
The function $f\prt{\s}$ parametrize the string world-sheet in the
bulk and moreover specifies the  initial angle for the spin.
%c3 The other half ($-L/2 \leq \s \leq 0$) of the string solution
%$z(\s)$ is to be obtained by symmetry.
The
%c1 initial
boundary conditions of the string are
%c1 together with the ones imposed by the symmetry of the space can be written as
\bea\lab{rotation}
&&z\prt{\pm \frac{L}{2}}=\infty~, \quad
\rho\prt{\pm\frac{L}{2},t}=\frac{L}{2}e^{-H t}~,\quad
\phi\prt{\pm \frac{L}{2},t}= \phi_\pm+\omega t~, \;
\eea
where the two endpoints of the string are antipodal in the equator, at
$\phi_+=0$ and at $\phi_-=\pi$.
%c3 move here
The parametrization \eq{spinning}, \eq{rotation}
describes a string on the equator of the spatial sphere, with
antipodal endpoints having an angular velocity $\omega$, and a
component of velocity transverse to the spinning motion and along the
axis that connects the endpoints, pointing inwards with measure
\eq{velocity}, just enough to counterbalance the time dependent
expansion of the dS boundary.
Below we will solve for the string solution for the region $0 \leq \s
\leq L/2$ subject to the boundary condition \eq{rotation} at the
endpoint $\s =L/2$. In addition we will require our solution to satisfy
\be
z'\prt{\s_0}=0~, \qquad z\prt{\s_0}=z_0~,
\ee
so that we can extend the solution to the other half $-L/2 \leq \s
\leq 0$. The constant $z_0$ is a free parameter that specifies the
coordinate of the tuning point of the string.

It turns out that the parametrization \eq{spinning} is a
consistent solution to the full
%c1
system
of equations of motion obtained by the NG
action
%c1 where
only if the function $f\prt{\s}$ is
%c1 found to be
given by
\be\label{functionf}
f\prt{\s}=\frac{\log(1-H^2\s^2)}{2 H}~.
\ee
Notice that the function \eq{functionf}  happens to be also part of
the coordinate transformation from the planar to the static
coordinates.
%c1 ; on that we elaborate further in the Appendix A.
Physically, it gives  a  non-trivial $\s$ dependence along the $U(1)$
angle which describes  a drag of the string profile in the bulk.

Having
specified the function $f\prt{\s}$, we
%c3 only
now need to solve the
equations of motion to determine $z\prt{\s}$ and obtain the string
profile in the bulk.
The on-shell action is time independent
\be%dg5
\frac{4\pi}{\sqrt{\l}\cT} S = \int d\s
\sqrt{\frac{1-\s^2\prt{H^2+\omega^2}}{1-H^2\s^2}\sinh^2 H
  z\prt{\sinh^2 H z+\prt{1-H^2\s^2}z'^2}}:=\int d\s \sqrt{D}~,
\ee
and depends explicitly on the worldsheet parameter $\s$.
Variation of the full  action gives only one independent equation of
motion and reads
\bea\nn
&&z''\sinh^2 H
z\prt{1-H^2\s^2}\prt{1-\prt{H^2+\omega^2}\s^2}
-z'^3\prt{H^2+\omega^2}\prt{1-H^2\s^2}^2\s \\\nn
&&-\frac{3}{2} z'^2\sinh 2 H z
\prt{1-H^2\s^2}\prt{1-\prt{H^2+\omega^2}\s^2} \nn\\
&& - z'\sinh^2 H
z\prt{\omega^2+2 H^2\prt{1-\s^2\prt{H^2+\omega^2}}}\s \nn \\
&&-\sinh^2 H z\sinh 2 H z \prt{1-\prt{H^2+\omega^2}\s^2}=0~.
\label{eomrotating}
\eea
The string carries energy and
%c4
angular
momentum defined by differentiating the
Lagrangian with respect to $\dot{t}$ and $\dot{\phi}$ and integrating
the densities along the length of the string
%c3 should not the limit be 0 and L/2?yes
\bea%dg5 typo
E_{Q\bar{Q}}&=&2\int_{0}^{L/2} d\s \; \frac{\prt{1- H^2\s^2} \sqrt{D}
  }{1-\s^2\prt{H^2+\omega^2}}~,\label{energyw0}\\
J_{Q\bar{Q}}&=&2\int_{0}^{L/2} d\s \;
\frac{-\omega \s^2}{1-\s^2(H^2+\omega^2)}\sqrt{D}~.
\label{momentumw0}
\eea
The
%c1
dependence on the parameter $\omega$ is
continuous and by switching it off $\omega=0$, the energy
\eq{energyw0} is equal to the static on shell action corresponding to
the energy of the bound state of the quark \eq{onshellaction1}, and
the angular momentum becomes null.
%c1
Due to the infinite length of the string both energy and
%c4
angular momentum are
infinite and a regularization is required.
The infinite terms that regularize the energy and the angular momentum
are given by \eq{Legendre} and applying it to our case we obtain
%c1 missing factor of 2 in the second term of both equation below?
\bea%dg5
%c3 no fraction! factor of 2%dg4
E&=& E_{Q\bar{Q}}-
2  \sinh H  z \frac{1-H^2\frac{L^2}{4}}{\sqrt{1-\frac{L^2}{4}(H^2+\omega^2)}}
\Bigg|_{z\rightarrow  \infty}~,
\label{energyw}
\\
J&=&J_{Q\bar{Q}}-
%c3 2 times
\frac{- L^2 \omega \sinh H z}{
2\sqrt{1-\frac{L^2}{4}(H^2+\omega^2)}
}
\Bigg|_{z\rightarrow  \infty}~.\label{momentumw}
\eea
For no rotation, the energy \eq{energyw} is equal to the one in
on shell action corresponding to
the energy of the bound state of the quark \eq{Etotalstatic}.

\subsection{The Spinning String Solutions}

%c1
The string profile is obtained first by solving the single equation
\eq{eomrotating} on the region $0 \leq \s \leq L/2$ and then extend to
the other half. That
%c3 that
this is possible is based on
the following observations:
 In the region close to the center of the sphere, the
string parametrized by  \eq{spinning}, \eq{rotation} does not have a
discontinuity along the $\phi$ coordinate, since the minimum point of
the string does not rotate. This can be checked by looking at the
equations of motion and noticing that in the full action the
derivatives $\partial_\s \phi^2 $ are multiplied by the term $ \s^2
\sinh^2 H z(\s)$. The function $z(\s)$ is even, with a minimum point
at $\s=0$, so its
%c1 first order
expansion around $\s =0$ is at least of second
order. Therefore in the resulting equations of motion the terms
involving derivatives of $\phi$ are all zero at $\s=0$ and the string
solution is smooth there too.
%c1 for $\s\pm \varepsilon$.
\begin{figure*}[!ht]
\begin{minipage}[ht]{0.5\textwidth}
\begin{flushleft}
\centerline{\includegraphics[width=78mm]{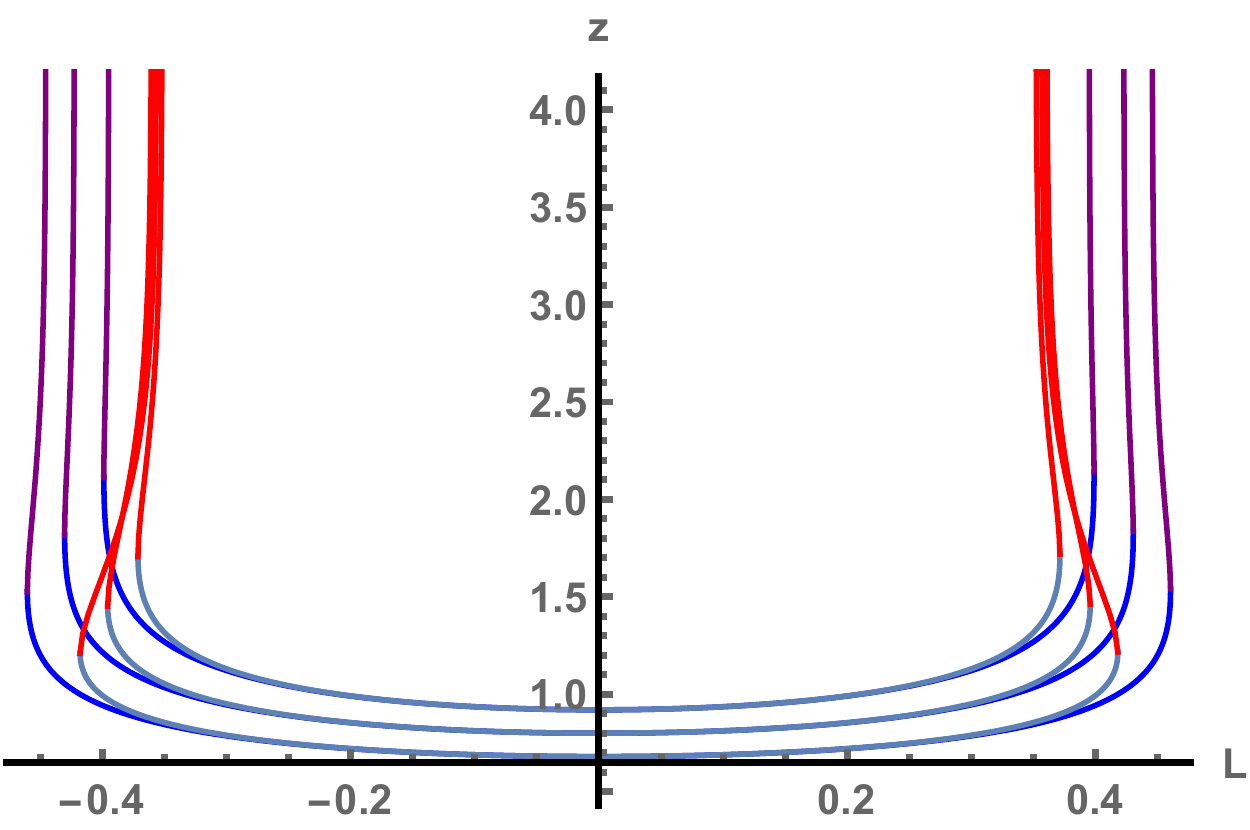}}
\caption{\small{The string world-sheet solution corresponding to a
    spinning meson($\omega=1$), compared to the static one for fixed
    bulk distance $z_0$. The blue-purple colored outer string
    solutions are the ones that correspond to the static string, while
    the inner red-cyan colored strings are the ones corresponding to
    the spinning state. Notice the spinning strings have less boundary
    distance between their endpoints, while the deformation of the
    U-shape string is enhanced due to rotation compared to the static
    configuration.}}
\label{figure:a6}
\end{flushleft}
\end{minipage}
\hspace{0.3cm}
\begin{minipage}[ht]{0.5\textwidth}
\begin{flushleft}
\centerline{\includegraphics[width=78mm ]{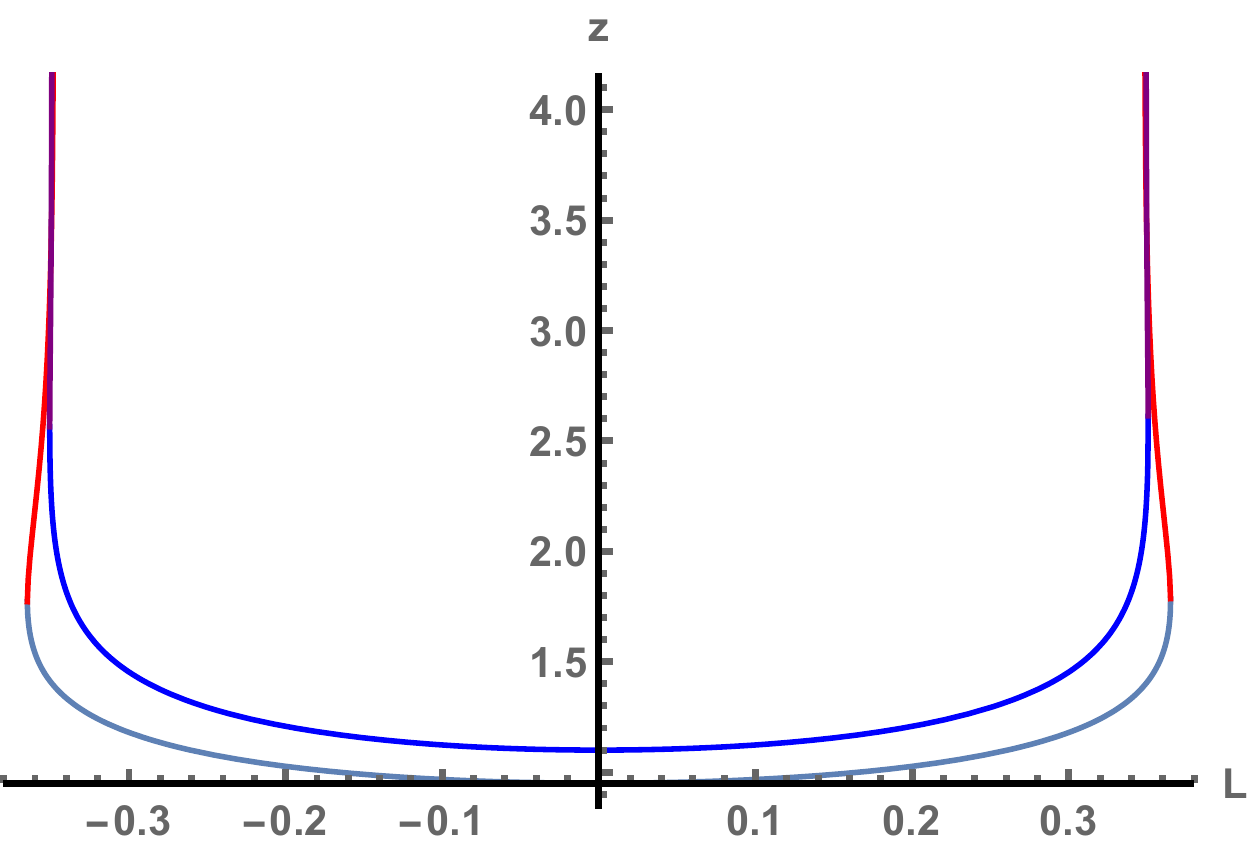}}
\caption{\small{Spinning and static string worldsheets with equal
    distance string endpoints. The spinning(red-cyan colored) string
    goes deeper in the bulk and is more deformed compared to the
    static one. The parameter $\omega$ is chosen equal to the unit.} }
\label{figure:a7}\vspace{2.4cm}
\end{flushleft}
\end{minipage}
\end{figure*}

%c1 The string solutions are given
Our obtained string profiles are presented
in the Figures \ref{figure:a6} and \ref{figure:a7}. The
deformation of the U-shaped string inside the bulk is enhanced by the
spin compared to the static case (Figure \ref{figure:a6}).
Moreover, spinning strings with
fixed turning point inter-quark distance $L$ correspond to surfaces
that go deeper to the bulk compared to the static strings (Figure
\ref{figure:a7}). This is naturally expected since rotation
%c1 of the state will aid
will tend to increase of the distance between the quarks, while
the string tension acts to the opposite direction. This is already a
hint that spinning strings dissociate easier than that static ones,
however for a rigorous evident we need to compute the energy of the
bound states.

\begin{figure*}[!ht]
\begin{minipage}[ht]{0.5\textwidth}
\begin{flushleft}
\centerline{\includegraphics[width=78mm]{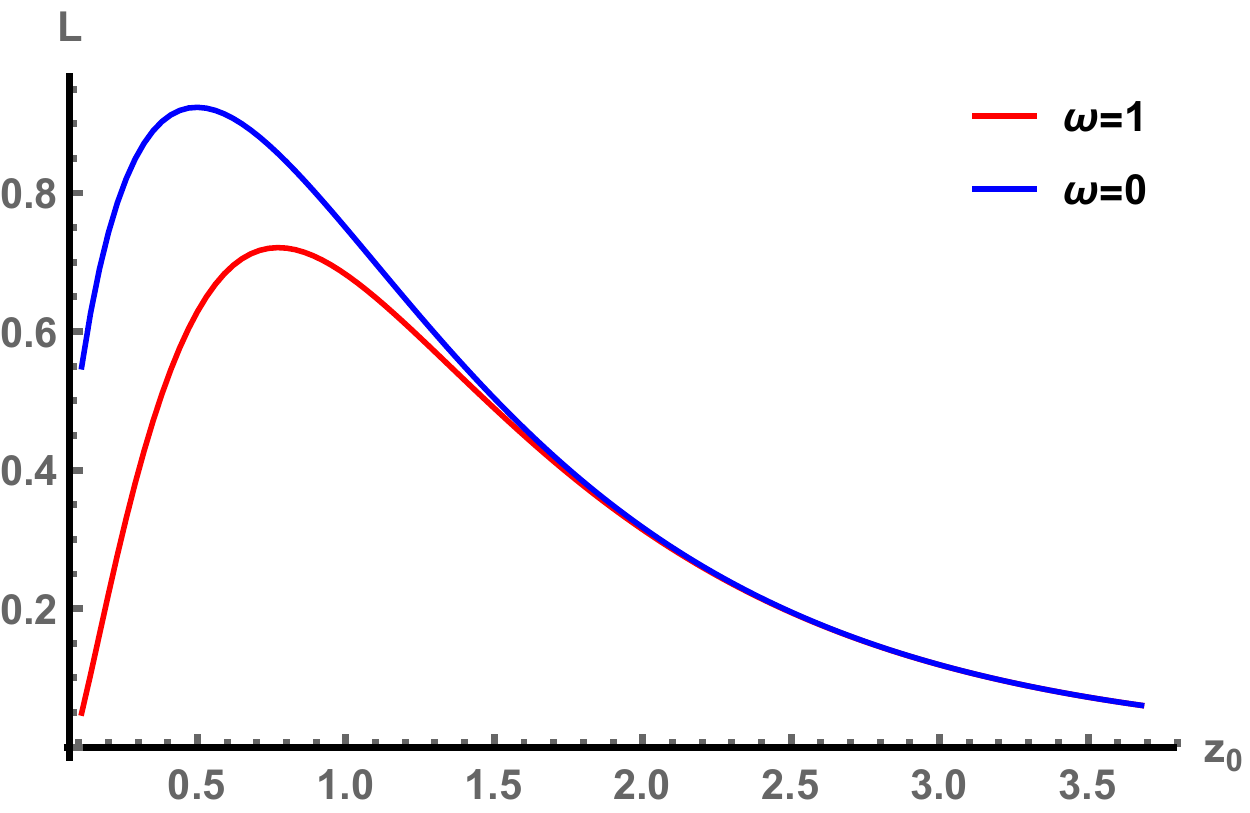}}
\caption{\small{The function $L(z_0)$ for fixed angular velocity. For each boundary condition, correspond two minimal surfaces. Increase of angular velocity, decreases the maximum length that the state can have.}}\vspace{1.1cm}
\label{figure:a8}
\end{flushleft}
\end{minipage}
\hspace{0.3cm}
\begin{minipage}[ht]{0.5\textwidth}
\begin{flushleft}
\centerline{\includegraphics[width=70mm ]{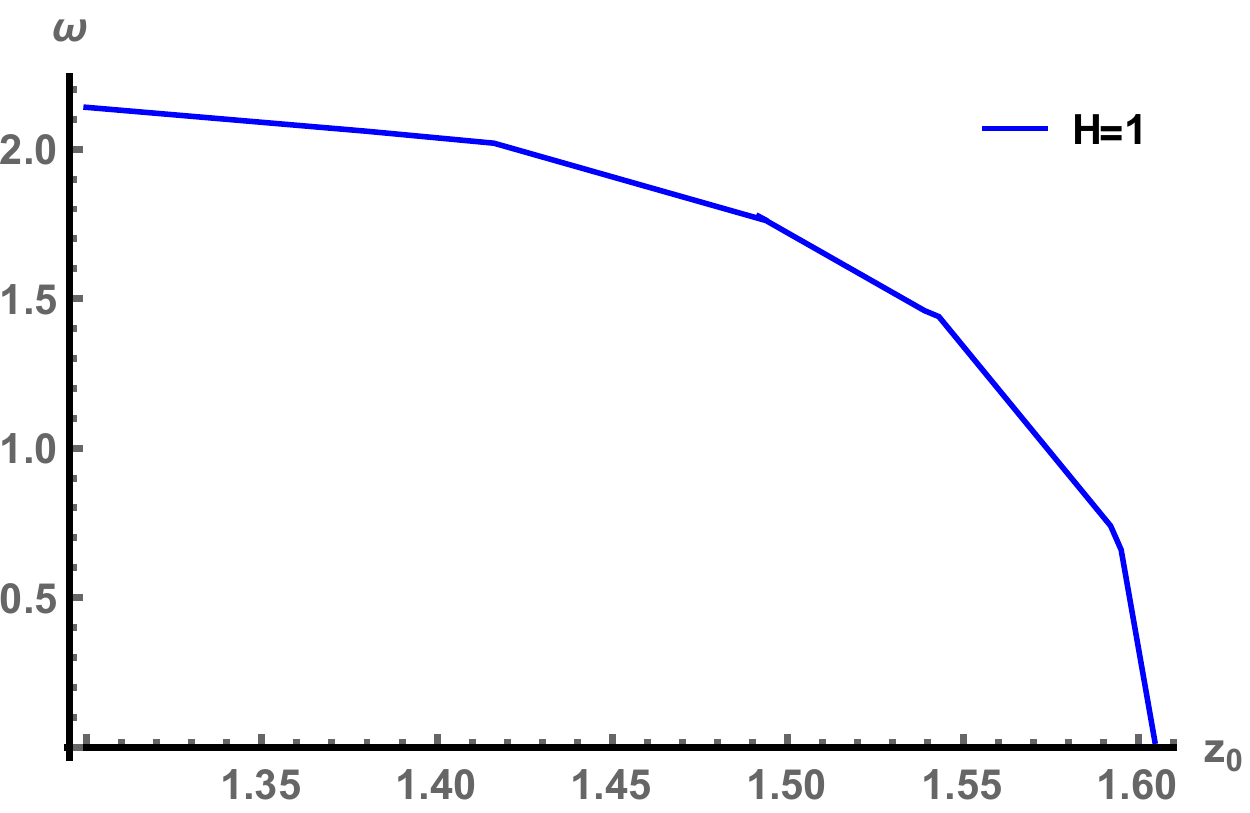}}
\caption{\small{The function $\omega(z_0)$ for a constant inter-quark
    distance at the boundary. Decrease of the angular velocity leads
    to surfaces that go deeper to the bulk  in order to preserve the
    length on the boundary. This explains naturally the finding of the
    Figure \ref{figure:a8}.} }
\label{figure:a9}
\end{flushleft}
\end{minipage}
\end{figure*}

In the case of spinning string we have two free parameters that the
energy of the state depends on. By fixing the spin of the string and
varying the string world-sheet endpoint we can obtain the function
$L(z_0)$. Then we can justify what we have already noticed by
comparing spinning and static strings: the first ones correspond to
larger inter-quark distances for the same turning point $z_0$, compared
to the latter ones. This is presented in Figure \ref{figure:a8}. When
we fix the length of the string world-sheet and increase the spin, we
obtain a function $\omega(z_0)$ where notice that the for higher
angular velocities the minimal surface goes deeper in the bulk in
order to preserve the invariant distance at the boundary (Figure
\ref{figure:a9}). This is in agreement with the previous observations
on the string profiles.

\begin{figure*}[!ht]
\begin{minipage}[ht]{0.5\textwidth}
\begin{flushleft}
\centerline{\includegraphics[width=78mm]{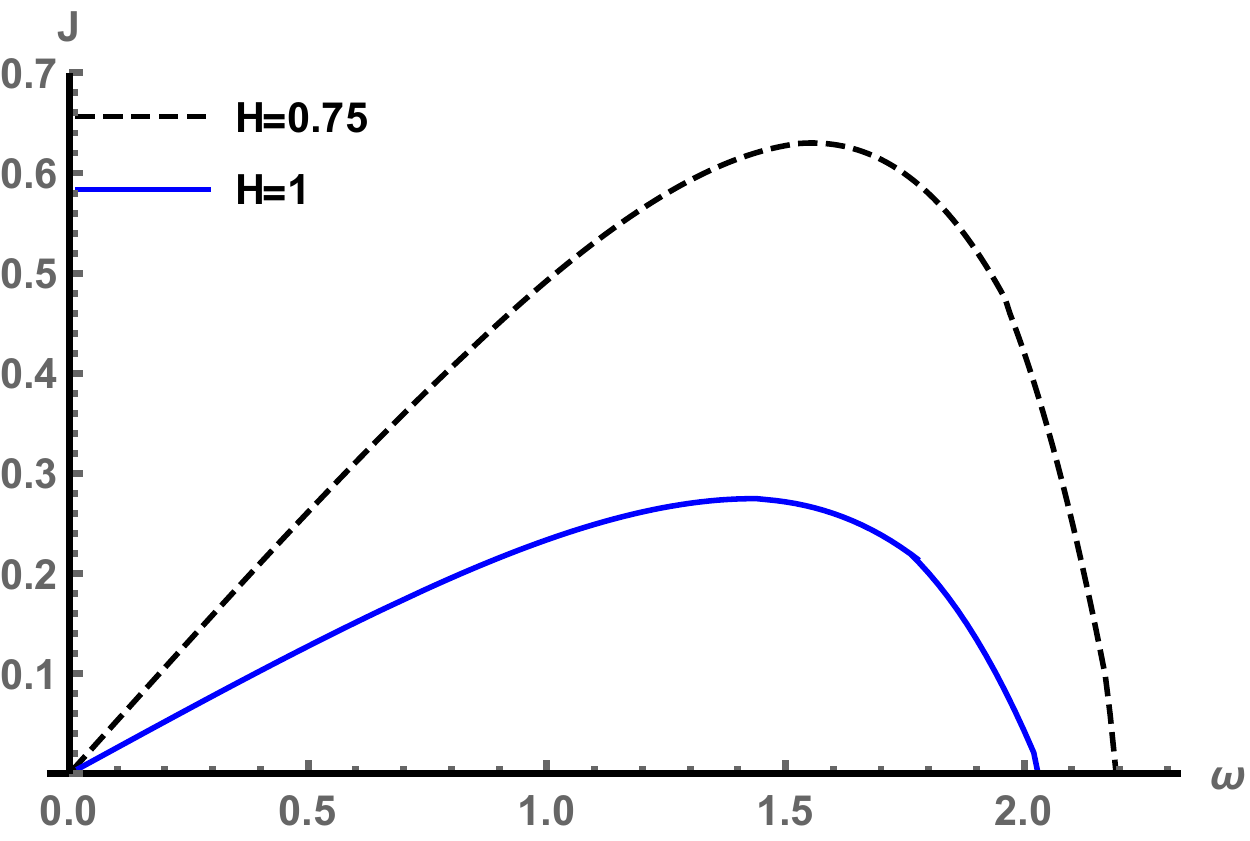}}
\caption{\small{The angular momentum in terms of the angular
    velocity. There is a maximum spin that the state can reach for a
    value of
%c3
$\omega_{\rm max}$.
Increase of the temperature allows larger
    values for the spin and steepest increase.}}\vspace{1.2cm}
\label{figure:a10}
\end{flushleft}
\end{minipage}
\hspace{0.3cm}
\begin{minipage}[ht]{0.5\textwidth}
\begin{flushleft}
\centerline{\includegraphics[width=72mm ]{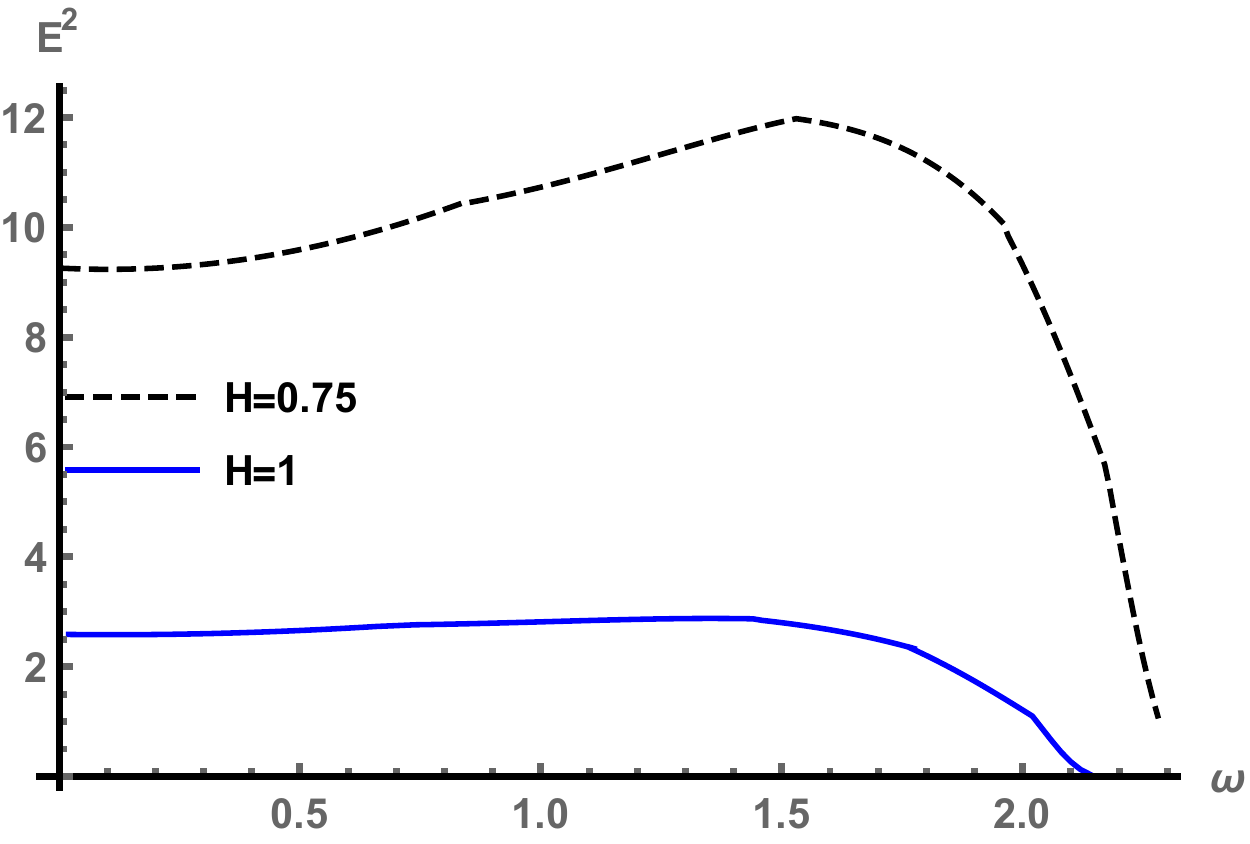}}
\caption{\small{The square of energy versus the angular velocity
    $E^2(\omega)$ for  constant inter-quark distance at the
    boundary. There is an increase of the energy until the angular
    velocity reaches the value
%c3
$\omega_{\rm max}$ and after that it
    decreases. The behavior resembles the angular momentum dependence
    on $\omega$, presented in Figure \ref{figure:a10}.} }
\label{figure:a11}\vspace{0cm}
\end{flushleft}
\end{minipage}
\end{figure*}

To observe the dissociation of the quark bound state in relation with
spin, we fix the length of the state
and the temperature,
and modify the
angular momentum. To keep a state in constant length as $\omega$
increases, we need a world-sheet that comes closer to the boundary. We
find that the
%c4
angular momentum of the bound state is increasing for increasing
$\omega$ until it reaches a maximum value for
%c3
$\omega=\omega_{\rm max}$,
where the decrease begins. For lower temperatures the magnitude of the
angular momentum is larger, and the heavy quark bound state can spin
faster. This is naturally expected since the string feels less
'friction' in lower temperatures (Figure \ref{figure:a10}).

\begin{figure}
\centerline{\includegraphics[width=78mm]{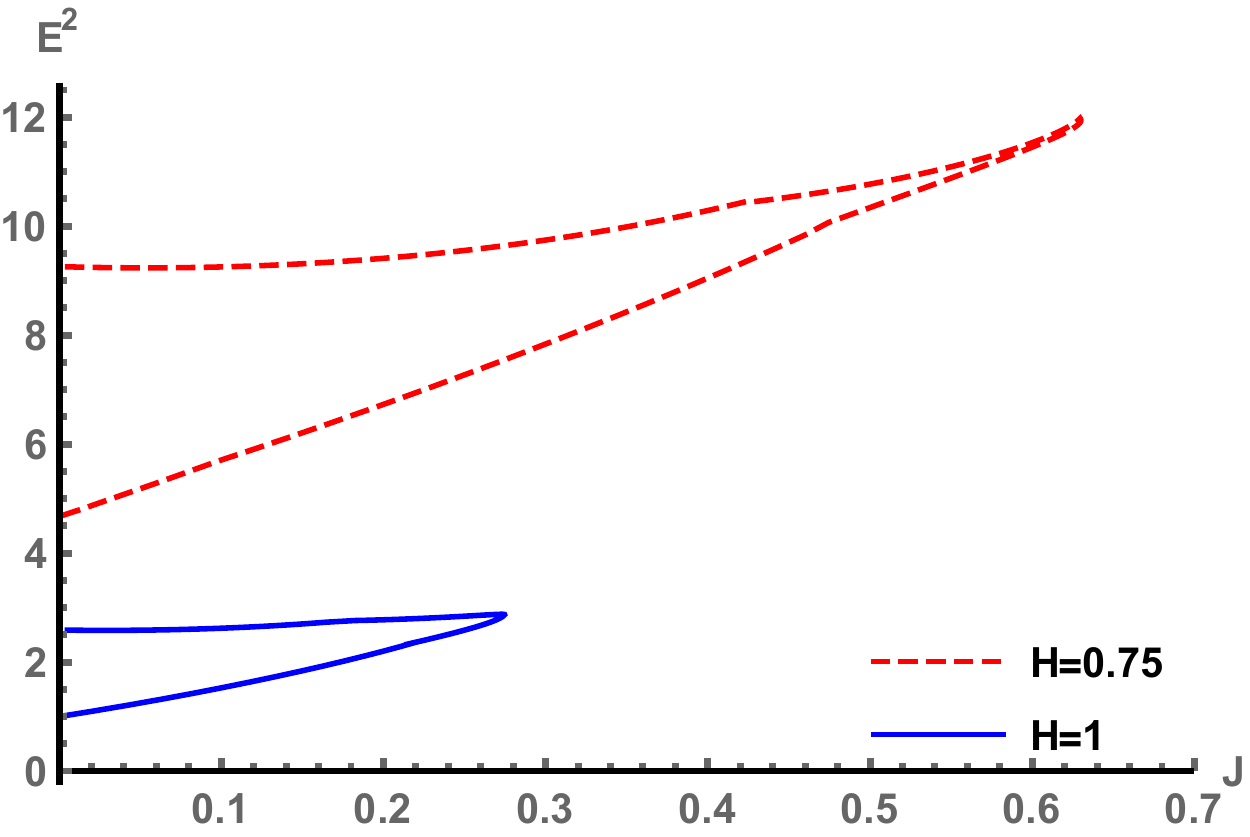}}
\caption{\small{The Energy versus the angular momentum for fixed
    boundary distance. Notice the maximum reachable spin for the bound
    state and the fact that for each value of angular momentum the
    state is allowed to have two energies.  The upper segment
    corresponds to large values of $\omega$ and is unstable and not
    energetically favorable. Increase of the temperature allow larger
    values for the spin, and decrease in the energies. The behavior of
    the bound state is like we have placed it in a finite temperature
    field theory %dg6
    in flat space.}}
\label{figure:a12}
\end{figure}

We find that the energy in terms of $\omega$ increases for increasing
angular velocity until it reaches a maximum value for
$\omega=\omega_{\rm max}$ (Figure \ref{figure:a11}). Lower temperatures
lead to higher energies $E(\omega)$ of the bound state with fixed
inter-quark distance.  The
%c3 maximal
maximum of the angular momentum and energy
occur
for the same value of angular velocity. Therefore the expression
$E^2(J)$ will have a cusp point at $J=J_{\rm max}$ indicating that a state
of fixed inter-quark distance can reach to a maximum spin.  Moreover,
for each value of spin $J<J_{\rm max}$ the state can be found with both
energies, the upper and lower segment in the $E^2(J)$ function (Figure
\ref{figure:a12}).%dg6 change #of fig
 The upper segment is for large values of $\omega$
and large energies and is not energetically preferable. The lower part
of the curve depicts the energy of a stable spinning state, and has a
continuous limit to the spin-less state.

The existence of a maximum energy and
%c3
angular momentum with respect to the
angular velocity can be explained by
looking at the behavior of the  minimal surface. We have
%c3 showed
shown that
for a fixed inter-quark distance and increasing the  spin,
the surface has to
extend deeper to the bulk in order to preserve the invariant distance
at the boundary. There is a critical point at
which effect of cosmological expansion and rotation becomes so big
that it is no longer possible to have a stable bound state.
%c3 where the extension of
% the surface to the bulk, reduces the total angular and energy momentum
% of the whole string and beyond this point the bound state becomes
% unstable.

To summarize our findings in this section, we conclude that the
spinning bound state on a dS CFT theory, has all
%c1 these
the characteristics of a spinning bound state in a finite temperature dual
field theory in flat spacetime. Therefore, it feels the heat bath in a
similar way as it would be in gravity dual theory with a black hole,
for example as in \cite{Peeters:2006iu}.
%c1 although the intermediate mechanisms are different.

\section{Conclusions and Discussions}

In
%c3 our
this paper we have examined the thermal properties of the dS field
theories in the Bunch-Davies vacuum. We have %dg6 studied 
provided a formalism for the study of  the entanglement
entropy of a strip in several dimensions. In the special case of
$AdS_3/dS_2$ sliced theory, the logarithmic entanglement entropy is
related to the geodesic in the bulk space, and due to conformal
invariance the result
%c3
is
expected to agree with the flat space field
theories. Nevertheless, we have studied
%c3 quite
analytically the entropy
and by using the geodesic approximation we have found the equal time
two point correlator and have reproduced the result obtained using the
bulk to boundary formalism in \cite{Chu:2016uwi}.

We have also
%c1 focused on
examined heavy quark probes placed in a strongly
coupled de Sitter universe. In the planar coordinates, the spatial
part of our metric expands with a rate of $\exp(H T)$. We  have
considered a pair of a heavy quark and a heavy anti-quark on the dS
boundary, where each quark has a constant speed pointing to each other
in order to counterbalance the expansion of the spacetime. The heavy
meson bound state has a constant invariant inter-quark distance which
corresponds to a time-translational invariant world-sheet. We have
discussed the numerical implications of the computation due to the
presence of cosmological
%c3 horizon
expansion in the bulk
and the way to  regularize the
infinite energy of the pair in holography. The meson bound state
%c1 has the common properties of
share many properties as the bound state in  thermal field theories.
%c1 and
In particular,
it does not exist beyond an inter-quark distance.

By examining the spinning heavy quark bound states our system gets one
more degree of freedom.
%c1 The velocity of each quark is decomposed into
% two components. A component pointing to the centre of
% the cycle where the quarks spin and counterbalances the expansion of
% the spacetime, while the other component is parallel to the cyclic
% trajectory and is responsible for the spinning motion. We parametrize
% the space in spherical coordinates and allow the meson to spin around
% the equator of the sphere with constant speed.
We observe that there
is a maximum angular momentum value that the spinning bound state can
exist. We compute the energy of the spinning meson, in terms of its
%c1
angular
momentum and conclude that the spinning string realizes the Hawking
temperature. It would be very interesting to develop a methodology for
other observables in the gravity dual of dS field theories,
especially the ones that their evaluation in flat thermal field
theories depend heavily on the presence of a black hole horizon, like
the jet quenching, and to examine how the generic formulas of
\cite{Giataganas:2012zy}
would be modified in the present setup.
%c3 are modified in the case of cosmological horizon.
Along these lines
we mention the interesting study of fluctuation and
dissipation in the de Sitter space \cite{Fischler:2014tka}.

It is worth noting that the effect of the
%c3 de Siter gravitational potential
cosmological expansion of the de Sitter factor persists in the bulk
and its effect on the string world-sheet
is evident. However the effect is different
compared to the strings placed in a black hole background
where the tidal gravitational attraction of the black hole tends to pull and
deforms the string
in the radial direction, while the cosmological expansion of the de
Sitter factor of the AdS space affects the string in the
longitudinal directions parallel to the
boundary.

%c1
In this paper we have consider de Sitter space written in planar or
conformal coordinates. Nevertheless we show that by a suitable choice
of ansatz for the string worldsheet, one can
eliminate from the effective system all the time dependence
consistently.  By doing that
we end up with time invariant system
%c1 surfaces and having to solve
and ordinary differential equations which are under a good control,
instead of the more involved partial differential
equations. Therefore,
%c1
the consideration in this paper may also provide some
%c1 our work also provides an initial step and
guidance towards the study of other observables in general
time dependent theories.

\appendix{\section{Strings in Static versus Planar
    Coordinates}} \label{appendix:a1}

Here we provide the coordinate transformation between the planar
coordinates of dS space
\be
ds^2=-d t^2+ e^{2 H t} (d r^2+ r^2 d\Omega)
\ee
and the static ones
\be
ds^2=-(1-H^2 \tilde{r}^2)d\tilde{t}^2+\frac{1}{1-H^2 \tilde{r}^2}
dr^2+\tilde{r}^2 d\Omega~.
\ee
We find the relation between the coordinate systems by identifying
\be
\tilde{r}=e^{H t} r~,
\ee
which leaves a differential equation to be solved for the time
coordinates to get the full transformation
\be
\tilde{t}=t-\frac{1}{2 H} \log(1-H^2 r^2 e^{2 H t})~,\qquad \tilde{r}=r e ^{H t}~.
\ee
One may use the coordinate transformation to map our string solutions
to the
static coordinates. By doing that to the static boundary bound state
\eq{wlansatz} we get on dS
\be
\tilde{t}= \t-\frac{1}{2 H} \log(1-H^2 \s)~,\qquad  \tilde{r}= \s~,
\ee
while the holographic coordinate is parametrized by another function $z(\s)$.
We notice that this string solution lead to a different action compared to the
solution obtained in \cite{Fischler:2014ama}, but to a similar
equation of motion. Moreover, the way we have parametrized the strings
in the planar coordinate system does not constrain us to have
to use radial coordinates with
a finite positive range,
%c2 $[0,r]$,
and we
%c3
are allowed to place
the string symmetrically, with respect to the origin.
%c2
We also note that with a coordinate transformation from the
 planar coordinates to the static ones,
one can bring the rotating string solution \eq{spinning}
to a form close to a boosted string in the static
coordinates.

\vskip7mm
%%%%%%%%%%%%%%%%%%%%%%%%%%%%%%%%%%%%%%%%%%%%%%%%%%%%%%%%%%%%%%%%%%%%%%%%%
\section*{Acknowledgements}
%%%%%%%%%%%%%%%%%%%%%%%%%%%%%%%%%%%%%%%%%%%%%%%%%%%%%%%%%%%%%%%%%%%%%%%%%
We would like to thank Koji Hashimoto and J. Pedraza for useful discussions.
%c3
We also thank the participants of the Academic Program ``Holography
and Topology of Quantum Matter'' (APCTP) for discussions
and the hospitality of APCTP during the final stage of this work.
This work is
supported in part by  the National Center of Theoretical Science
(NCTS) and the
grants  101-2112-M-007-021-MY3 and 104-2112-M-007 -001 -MY3 of the
Ministry of Science and
Technology of Taiwan.

%%%%%%%%%%%%%%%%%%%%%%%%%%%%%%%%%%%%%%%%%%%%%%%%%%%
%\begin{thebibliography}{100}
\bibliographystyle{JHEP}

%\bibliography{botany}
%\end{thebibliography}
\end{document}